\documentclass[10pt,journal]{IEEEtran}

\usepackage[english]{babel}
\usepackage[T1]{fontenc}
\usepackage[utf8]{inputenc}
\usepackage{xargs}

\usepackage{amssymb}
\usepackage{mathtools}
\usepackage{nccmath}

\usepackage{amsmath}
\usepackage{bbm}
\usepackage{bm}

\usepackage{stmaryrd}
\usepackage{stackengine}
\stackMath

\usepackage{graphicx}
\usepackage[dvipsnames]{xcolor}
\usepackage{tabularx}

\ifCLASSOPTIONcompsoc
\usepackage[caption=false,font=normalsize,labelfont=sf,textfont=sf]{subfig}
\else
\usepackage[caption=false,font=footnotesize]{subfig}
\fi

\usepackage[ruled,norelsize]{algorithm2e}
\makeatletter
\newcommand{\removelatexerror}{\let\@latex@error\@gobble}
\makeatother

\usepackage{etoolbox}
\usepackage{tikz}
\usetikzlibrary{tikzmark}
\usetikzlibrary{calc}

\usepackage{hyperref}

\usepackage{booktabs}
\usepackage{colortbl}
\usepackage{threeparttable}

\usepackage{siunitx}
\DeclareSIUnit{\au}{a.u.}

\usepackage{relsize}

\usepackage{xspace}
\DeclarePairedDelimiter{\paren}{\lparen}{\rparen}
\DeclarePairedDelimiter{\parenEl}{[}{]}
\DeclarePairedDelimiter{\set}{\lbrace}{\rbrace}

\newcommand{\interval}[4]{\mathopen{#1}#2 \mathclose{}\mathpunct{}, #3\mathclose{#4}}
\newcommand{\intervalcc}[2]{\interval{[}{#1}{#2}{]}}

\newcommand{\intervalco}[2]{\interval{[}{#1}{#2}{)}}

\newcommand\funop[1]{\mathop{{}#1}}

\DeclarePairedDelimiter{\inner}{\langle}{\rangle}

\newcommand\PS{{\mathrm{ps}}}
\newcommand\MT{{\mathrm{m}}}
\newcommand\C{{\mathrm{c}}}

\newcommand\BG{\mathrm{bg}}

\newcommand{\pmSmooth}{\ensuremath{{\mathrm{pm}1}}\xspace}
\newcommand{\pmSparse}{\ensuremath{{\mathrm{pm}2}}\xspace}
\newcommand\MAP{\mathrm{MAP}}

\newcommand\len{\ell}
\newcommand\dlen{\diff\len}

\renewcommand{\vec}[1]{\bm{#1}}
\newcommand*\x{\ensuremath{\vec{x}}\xspace}

\newcommand*\y{\ensuremath{\vec{y}}\xspace}

\newcommand*\dir{\ensuremath{\vec{d}}\xspace}

\newcommand\onesVec{\vec{1}}
\newcommand\zerosVec{\vec{0}}

\renewcommand\matrix[1]{\bm{#1}}
\newcommand{\transpose}[1]{#1^T}
\newcommand{\inverse}[1]{\paren*{{#1}}^{-1}}
\DeclareMathOperator*{\diag}{diag}

\DeclarePairedDelimiter{\Lone}{\lVert}{\rVert_1}
\DeclarePairedDelimiterXPP\Ltwo[1]{}{\lVert}{\rVert}{_2}{#1}
\DeclarePairedDelimiterXPP\Linf[1]{}{\lVert}{\rVert}{_\infty}{#1}

\DeclarePairedDelimiter{\abs}{|}{|}

\newcommand*\R{\mathbb{R}}
\newcommand*\Rpos{\R_{+}}
\newcommand*\Z{\mathbb{Z}}

\newcommand{\diff}{\mathop{\mathrm{{}d}}\mathopen{}}
\newcommand*\dx{\diff \x}
\newcommand*\dt{\diff \t}
\newcommand*\dy{\diff \y}

\newcommand*\dirac{\delta}
\newcommand\Poisson{\mathrm{Poisson}}

\DeclareMathOperator{\vectorise}{vec}
\newcommand\ConvSymbol{*}
\newcommand\conv{\mathbin{\ConvSymbol}}
\newcommand*{\defeq}{\mathrel{\coloneqq}}
\newcommand*{\eqdef}{\mathrel{\eqqcolon}}

\newcommand{\indicator}{\ensuremath{\mathbbm{1}}}
\newcommand{\1}[1]{\funop{\indicator_{\!#1}}}
\newcommand{\indicatorOptim}{\ensuremath{\iota}}
\newcommand\inidicatorSet{\mathcal{S}}

\newcommand\expe{\mathrm{e}}
\newcommand\psnr{\ensuremath{\mathrm{PSNR}}\xspace}
\newcommand\nll{\mathrm{nll}}
\newcommand\nllNormed[1][\PX]{\mathrm{nll}_{#1}}

\newcommand\ceil[1]{\mathrm{ceil}\paren{#1}}

\newcommand\params{{\vec{\theta}}}
\newcommand\paramsHat{\widehat{\params}}
\newcommand\paramSet{\vec{\Theta}}

\newcommand\px{\mathcal{P}}
\newcommand\iPx{\ensuremath{j}\xspace}
\newcommand\iPxSet{\mathcal{J}}

\newcommand\pxArea{\abs{\px}}
\newcommand*\discPos{\x_\iPx}

\renewcommand\t{\ensuremath{t}\xspace}
\newcommand{\timeSet}{\ensuremath{\mathcal{T}}\xspace}
\newcommand*\discTime{\t_\mathsmaller{\!\!\:\iFrame}}

\newcommand\iObject{\ensuremath{l}\xspace}
\newcommand\objectL{\mathcal{L}}
\newcommand\objMeas[1][\t]{\ensuremath{\phi_{#1}}\xspace}
\newcommand\photomMeas[1][]{\ensuremath{\phi}_{#1}\xspace}
\newcommand\intensSymbol{\varphi}
\newcommand\intens[1][\t]{\intensSymbol_{#1}}
\newcommand\intensInt[1][\t]{\integrated{\intensSymbol}_{#1}}
\newcommand\phFlux[1][]{\Phi^{#1}}

\newcommand\integrated[1]{\overline{#1}}
\newcommand\phFluxInt[1][]{\integrated{\Phi}^{#1}}

\newcommand\bgPhi[1][\t]{\intens[#1]^{\BG}}
\newcommand\bgPhiInt[1][\t]{\intensInt[#1]^{\BG}}

\newcommand\manifold[1][\t]{\ensuremath{\mathcal{M}_{#1}}\xspace}

\newcommand\psPos[1][\t]{\x_{#1}^{\PS}}

\newcommand\VS{\mathcal{VS}}
\newcommand\iVS{\ensuremath{s}\xspace}
\newcommand\nVS[1][\t]{{n^{\!\mathrm{v}}_{#1}}}
\newcommand\weightVS[1][\t]{w_{#1}}
\newcommand\weightVSInt[1][\t]{\integrated{w}_{#1}}
\newcommand\weightVSVecInt[1][\t]{\integrated{\vec{w}}_{#1}}
\newcommand\weightQ[1][\t]{w_{#1}^{\mathrm{q}}}

\newcommand\curve[1][\t]{\ensuremath{\mathcal{C}_{\!\!\;#1}}\xspace}
\newcommand\RiemannianMetric[1][\t]{\ensuremath{g_{#1}}}

\newcommand\paramSpace[1][\t]{\mathcal{D}_{#1}}
\newcommand\paramSpaceHat[1][\t]{\widehat{\mathcal{D}}_{#1}}

\newcommand\curveLen[1][\t]{\Lambda_{#1}}
\newcommandx\curveDir[2][1=, 2=\t]{\dir_{#2}^{#1}}

\newcommand\mtLen[1][\t]{\curveLen[\MT][#1]}
\newcommand\mtPhi[1][\t]{\intens[#1]}
\newcommand\mtPhiMAP[1][\t]{\widehat{\intensSymbol}_{#1}^{\MAP}}
\newcommand\mtPhiVec[1][\t]{\vec{\intensSymbol}_{#1}}
\newcommand\mtPhiVecMAP[1][\t]{\widehat{\vec{\intensSymbol}}^{\MAP}_{#1}}

\newcommand\mapParamToObj[1][\t]{\vec{\sigma}_{#1}}
\newcommand\mapParamToObjHat[1][\!\t]{\widehat{\vec{\sigma}}_{#1}}

\newcommand\mapParamToObjInv{\vec{\sigma}_t^{-1}}
\newcommand\mapPxToImg{\nu}
\newcommand\mapImgToPx{\mapPxToImg^{-1}}

\newcommand\obsPhCount[1][\iFrame\iPx]{\ensuremath{N^\textrm{photon}_{#1}}\xspace}
\newcommand\obsPhCountVec[1][\iFrame]{\ensuremath{\vec{n}_{#1}}\xspace}
\newcommand\expPhCount[1][\iFrame\iPx]{\ensuremath{\mu_{#1}}\xspace}
\newcommand\expPhCountVec[1][\iFrame]{\ensuremath{\vec{\mu}_{#1}}\xspace}

\newcommand\obsGVCount[1][\iFrame\iPx]{\ensuremath{N^\textrm{grey}_{#1}}\xspace}
\newcommand\obsGVMat[1][\t]{\matrix{Z}_{#1}}

\newcommand\nFrames{{n^{\mathrm{f}}}}
\newcommand\nSlices{{n^{\!\!\;\mathrm{s}}}}  
\newcommand\nHeight{{n^{\!\!\;\mathrm{h}}}}  
\newcommand\nWidth{{n^{\!\!\;\mathrm{w}}}}  
\newcommand\nPx{\ensuremath{{n^{\mathrm{p}}}}\xspace}

\newcommand\iFrame{k}
\newcommand*\frameSet{\mathcal{F}}
\newcommand\tA{\ensuremath{\t_a}\xspace}

\newcommandx\tAin[1][1=t_{\iFrame}]{\ensuremath{\intervalcc{#1}{#1 + \tA}}\xspace}

\newcommand\psfFun{\kappa}
\newcommand\psfVec[1][\t]{\vec{\psfFun}_{#1}}

\newcommand\psfMat[1][\t]{\matrix{K}_{#1}}

\newcommand\SG{\mathrm{sg}}
\newcommand\psfFunSG{\psfFun^{\SG}}
\newcommand\constSG{C^{\SG}}

\newcommand\wavelength{\lambda}
\newcommand\quantEfficiency{q_{\wavelength}}
\newcommand\mGain{M}
\newcommand\ADU{f}
\newcommand\cameraBias{b}

\newcommand\whitenedOp[1][\t]{L_{#1}}
\newcommand\whitenedOpMat[1][\t]{\matrix{L}_{#1}}
\newcommand\innov[1][\t]{w_{#1}}
\newcommand\innovVec[1][\t]{\vec{u}_{#1}}
\newcommand\potential{\varpi_{\mathsmaller U}}
\newcommand\potentialNormed[1][\t]{\varpi_{#1,\mathsmaller U}}

\newcommand\kymoNN{\widehat{\mathcal{K}}^{\textrm{NN}}}
\newcommand\kymoMAP{\widehat{\mathcal{K}}^{\textrm{MAP}}}
\newcommand\kymoSpace{\mathcal{K}}

\newcommand\binSize{\Delta_{\len}}

\newcommand{\iBase}{\ensuremath{p}\xspace}
\newcommand{\nBases}[1][\t]{{n^{\!\mathrm{b}}_{#1}}}
\newcommand*{\basisFun}{\beta}
\newcommand*{\basisVec}[1][\t]{\vec{\basisFun}_{#1}}
\newcommand\basisMat[1][\t]{\matrix{B}_{#1}}
\newcommand\kymoToImageMat[1][\t]{\vec{M}_{#1}}

\newcommand\imageDom{\Omega}

\newcommand\prob{\mathbb{P}}
\newcommand\probInnov{\prob_U}

\DeclareMathOperator{\shrink}{shrink}
\DeclareMathOperator{\sign}{sign}

\newcommand\energyMAP{\mathcal{E}^\MAP}

\newcommand\eyeMat{\vec{I}}

\newcommand\reg{\eta}

\newcommand\step{\gamma}

\newcommand\iIter{i}

\newcommand\operator{\vec{O}}

\DeclareMathOperator{\Prox}{Prox}
\DeclareMathOperator{\Proj}{Proj}

\renewcommand{\b}{\vec{b}}
\newcommand{\bc}{\vec{b}^i}
\newcommand{\bn}{\vec{b}^{i + 1}}
\newcommand\w{\vec{w}}
\newcommand{\wc}{\w^i}
\newcommand{\wn}{\w^{i + 1}}

\newcommand\mtPhiLS{\vec{\varphi}}

\newcommand\mtPhiLSn{\mtPhiLS^{\iIter + 1}}
\newcommand\mtPhiNLL{\vec{w}_{1}}

\newcommand\mtPhiNLLn{\mtPhiNLL^{\iIter + 1}}
\newcommand\bMtPhiNLL{\vec{b}_{1}}
\newcommand\bMtPhiNLLc{\bMtPhiNLL^{\iIter}}

\newcommand\mtPhiT{\vec{w}_{2}}

\newcommand\mtPhiTn{\mtPhiT^{\iIter + 1}}
\newcommand\bMtPhiT{\vec{b}_{2}}
\newcommand\bMtPhiTc{\bMtPhiT^{\iIter}}

\newcommand\mtPhiP{\vec{w}_{3}}

\newcommand\mtPhiPn{\mtPhiP^{\iIter + 1}}
\newcommand\bMtPhiP{\vec{b}_{3}}
\newcommand\bMtPhiPc{\bMtPhiP^{\iIter}}

\makeatletter
\newcommand\ensuresingleperiod{\@ifnextchar.{}{.\@\xspace}}
\makeatother

\newcommand\ie{\textit{i.e.}\xspace}
\newcommand\eg{\textit{e.g.}\xspace} 

\newcommand\aposteriori{\textit{a posteriori}\xspace}

\newcommand\invitro{\textit{in vitro}\xspace}
\newcommand\invivo{\textit{in vivo}\xspace}

\newcommand\SCerevisiae{\textit{S. cerevisiae}\xspace}

\renewcommand\refeq[1]{\eqref{#1}}
\newcommand\reffig[1]{Fig.~\ref{#1}}
\newcommand\refsubfig[1]{\protect\subref{#1}}
\newcommand\refalg[1]{Algorithm~\ref{#1}}
\newcommand\reftable[1]{Table~\ref{#1}}
\newcommand\refsection[1]{Section~\ref{#1}}

\usepackage{amsthm}
\newtheorem{result}{Result}
\newtheorem*{result*}{Result}

\title{A Bayesian framework for the analog reconstruction of kymographs from
  fluorescence microscopy data}

\author{Denis~K.~Samuylov, Gábor~Székely, and~Grégory~Paul}

\begin{document}
\maketitle

\begin{abstract}
%
Kymographs are widely used to represent and analyse spatio-temporal dynamics
of fluorescence markers along curvilinear biological compartments.
These objects have a singular geometry, thus kymograph reconstruction is
inherently an analog image processing task.
However, the existing approaches are essentially digital: the kymograph
photometry is sampled directly from the time-lapse images.
As a result, such kymographs rely on raw image data that suffer from the
degradations entailed by the image formation process and the spatio-temporal
resolution of the imaging setup.
In this work, we address these limitations and introduce a well-grounded
Bayesian framework for the analog reconstruction of kymographs.
To handle the movement of the object, we introduce an intrinsic description of
kymographs using differential geometry: a kymograph is a photometry defined on
a parameter space that is embedded in physical space by a time-varying map
that follows the object geometry.
We model the kymograph photometry as a Lévy innovation process, a flexible
class of non-parametric signal priors.
We account for the image formation process using the virtual microscope
framework.
We formulate a computationally tractable representation of the associated
maximum \aposteriori problem and solve it using a class of efficient and
modular algorithms based on the alternating split Bregman.
We assess the performance of our Bayesian framework on synthetic data and
apply it to reconstruct the fluorescence dynamics along microtubules \invivo
in the budding yeast \SCerevisiae.
We demonstrate that our framework allows revealing patterns from single
time-lapse data that are invisible on standard digital kymographs.

\end{abstract}

\begin{IEEEkeywords}
  Analog reconstruction, Bayesian modelling, deconvolution, fluorescence
  microscopy, inverse problems, Lévy innovation processes, model-based image
  processing, operator splitting, super-resolution, virtual microscope
\end{IEEEkeywords}


\section{Introduction}

%
\IEEEPARstart{F}{luorescence} microscopy is a powerful tool for studying the
dynamics of biological processes.
In many cases, these processes take place in specific compartments. As
geometries, these compartments represent different levels of spatial
restriction: inside a volume (\eg cytoplasm), on a surface (\eg membranes), or
along a curvilinear object (\eg axons, microtubules, actin filaments). In this
paper, we focus on processes restricted to curvilinear geometries.
A standard representation for characterising the spatio-temporal dynamics of
such processes is a \emph{kymograph}~\cite{Hinz1999}: a two-dimensional
representation of the signal along the curvilinear object (\eg displayed as the
ordinate) varying in time (\eg displayed as the abscissa).
Kymographs are custom in biology to study regulatory mechanisms coordinating
spatio-temporal protein interactions (see \eg
\cite{Bieling2007,Racine2007,Welzel2009,Roberts2014}). Thus, there is a high
interest in developing computational tools to build and analyse
kymographs~\cite{Chiba2014,Mangeol2016,Chaphalkar2016,Neumann2017}.

%
Reconstructing kymographs from a sequence of images is a challenging image
processing task.
It requires jointly estimating \emph{the geometry and the photometry} of a
curvilinear object.
When the objects move during the time-lapse imaging, the complexity of the task
increases as in addition it requires solving a \emph{tracking problem}.
Reconstructing a kymograph is an \emph{inherently analog image processing task}
because the typical diameter of such biological objects is smaller than the
pixel size, and not aligned with the pixel grid geometry.
To the best of our knowledge, the existing kymograph reconstruction frameworks
have been mostly digital in nature: the kymograph is reconstructed by directly
sampling the image data, with an optional pre- or post-processing. In addition,
the geometry and the photometry are estimated sequentially.

A standard approach to reconstruct kymographs consists of mainly two steps:
estimating the curvilinear object geometry and sampling the grey values along
this geometry.
When the image data are acquired at a single focal plane, the geometry is
outlined
manually~\cite{Racine2007,Zhang2011,Chiba2014,Mangeol2016,Chaphalkar2016,Neumann2017},
semi-automatically~\cite{Racine2007,Mukherjee2010}, or
automatically~\cite{Racine2007} on the average or on the maximum projection of
the time-lapse data.
When the image data are acquired at multiple focal planes, the maximum
projection is first applied along the axial
direction~\cite{Mennella2005,Racine2007}. The axial coordinate is either
discarded~\cite{Mennella2005} or obtained from a manual annotation of the
curvilinear object geometry on a cross-sectional image between the time-lapse
images projected in time and the outlined two-dimensional
geometry~\cite{Racine2007}.
It is common to dilate the estimated geometry and orthogonally project the
enclosed pixel onto the estimated curve by computing the maximum
projection~\cite{Kner2009,Mukherjee2010,Pereira2010,Chetta2011,Chiba2014}.
It makes the reconstructed kymographs robust to small errors in the estimated
geometry and its variations during time-lapse imaging. It also allows observing
the intensity variations within a region of interest~\cite{Kner2009}.
However, the kymograph reconstruction will fail in case of high object
displacement due to drifting or to the underlying object
dynamics~\cite{Chetta2011}.
To address the problem of reconstructing kymographs, it has been suggested to
track their geometry using semi-~\cite{Chetta2011} or
fully-automated~\cite{Pereira2010} approaches.
To facilitate the kymograph processing (\eg estimating trajectory, direction,
velocity of vesicles that moves along the curvilinear geometries), it has been
suggested to denoise~\cite{Chiba2014} and deconvolve the original
time-lapse~\cite{Racine2007} as a pre-processing step, or to apply digital
filters to the kymographs as a post-processing
step~\cite{Racine2007,Mukherjee2010,Chetta2011}.
In~\cite{Kner2009}, a super-resolution microscopy technique
(structured-illumination microscopy operated in an total internal reflection
fluorescence mode) yields high-resolution images, and hence high-resolution
kymographs.
%

The aforementioned digital techniques reconstruct kymographs by sampling or
averaging image grey values that suffer from two sources of degradation:
\emph{distortions} due to light emission, propagation through the environment
and conversion into grey values;
\emph{spatio-temporal resolution limit} entailed by the pixel grid and the
acquisition duration.
Therefore, these approaches implicitly assume that these degradations are
negligible. This assumption is reasonable for objects that are large and
immobile compared to the spatio-temporal resolution of the imaging setup.
Those factors set limits on the spatio-temporal scales that can be directly
resolved, decrease the quality of the kymographs and complicate their subsequent
analysis~\cite{Yuan2012,Zhang2011}.
%

To address these limitations, we establish a minimal Bayesian framework that
allows formulating inverse problems for the analog reconstruction of kymographs
in fluorescence microscopy.
We assume that the geometry is given as the solution of another inverse problem
that is not the focus of this paper.
The main challenge addressed in this work is to formulate and solve an
\emph{analog inverse problem} to reconstruct the kymograph before degradation.




\section{Motivation: A Super-resolution perspective}

Super-resolution is an inverse problem characterised by a reconstruction space
that is different from the sampling space. Two archetypal super-resolution
problems are \emph{zooming} and \emph{source localisation}: zooming aims at
reconstructing an image on a grid finer than the sampled data; source
localisation amounts to reconstructing positions off the sampling grid.

Following~\cite{Samuylov2015}, imaged objects are characterised by geometry and
photometry. Informally, geometry describes the light sources location, and
photometry their intensity value. In zooming, the main focus is reconstructing
photometry, whereas geometry is either given or a nuisance parameter. In
contrast, in source localisation, the main interest is geometry, and photometry
is treated as a nuisance parameter. In the literature, kymograph reconstruction
assumes a given geometry and concentrate on reconstructing photometry.
Therefore, the problem we address in this work is similar to the zooming
problem, but on a curve.

\subsubsection{The geometry and photometry of deconvolution}
In this section we revisit the convolution linear inverse problem from the
perspective of geometry and photometry. The mean intensity $\vec{\mu}$ generated
by a distribution of $n_s$ point sources sampled on a grid of $n_p$ pixels is
modelled as a linear equation:
\begin{equation}
  \label{eq:linear-convolution}
  \def\stackalignment{c}
  \def\stacktype{L}
  \setstackgap{L}{.7\normalbaselineskip}
  \stackunder{\vec{\mu}}{\mathsmaller{n_p \times 1}}
  =
  \stackunder{\vec{\mu}_{\textrm{bg}}}{\mathsmaller{n_p \times 1}}
  +
  \stackunder{\vec{M}}{\mathsmaller{n_p \times n_s}}\,
  \stackunder{\vec{\varphi}}{\mathsmaller{n_s \times 1}}\ ,
\end{equation}
where $\vec{\mu}_{\textrm{bg}}$ is an offset due to background, and $\vec{M}$ is
a matrix representing both convolution and sampling (integration in space and
time). Each row of $\vec{M}$ corresponds to a pixel grid location, and each
column to a point source location. When these two sets coincide, we obtain the
custom convolution matrix~\cite{Hansen2006}. In addition, each column of
$\vec{M}$ holds the integrated/sampled point spread function (PSF) shifted at
each point source location. The vector $\vec{\varphi}$ hence represents the
intensity of each point source. This straightforward identification reveals that
geometry is \emph{implicitly} encoded in the column space of the matrix
$\vec{M}$, whereas photometry is \emph{explicitly} encoded in $\vec{\varphi}$.
This insight is crucial to motivate the approach we develop for solving the
kymograph reconstruction problem.

\subsubsection{Interplay between geometry and photometry in the conditioning of
  deconvolution}
\label{sec:interpl-betw-geom}
The condition number of $\vec{M}$, denoted
$\kappa(\vec{M}) \in \intervalcc{1}{\infty}$, quantifies the sensitivity of the
solutions of the linear inverse problem~\eqref{eq:linear-convolution} to
perturbations in $\vec{\mu}-\vec{\mu}_\textrm{bg}$,
see~\cite{Golub2012,Vogel2002}. The reciprocal condition number (RCN), denoted
$1/\kappa \in \intervalcc{0}{1}$, is a \emph{scaled distance to the nearest
  ill-posed problem} \cite{Demko1986,Demmel1987}. It assumes the value 0 for
ill-posed problem, and the closer to 0, the more ill-conditioned.

We have shown previously that the column space of $\vec{M}$ is related to the
localisation of the point sources. Therefore, the conditioning of the
deconvolution problem~\eqref{eq:linear-convolution} is related to geometry. To
quantify this relationship, we need another insight about condition numbers:
they quantify collinearity in column space.

\begin{result}{RCN for the linear deconvolution
    problem~\eqref{eq:linear-convolution}.}
  \label{result:rcn}
  The linear deconvolution problem is ill-posed whenever at least two point
  sources occupy exactly the same position. In such a case,
  equation~\eqref{eq:linear-convolution} admits an infinite number of solutions.

  The deconvolution problem is well-posed when all point sources occupy
  different positions. The RCN decreases with the total overlap of the PSF
  kernels. For compactly supported PSFs, the maximum RCN is reached when none of
  the support overlap. In such a case, the RCN equals the ratio between the
  smallest and the largest norms among the column vectors of $\vec{M}$, \ie
  $\min_k \Ltwo{\vec{M}_{\cdot k}}/\max_k \Ltwo{\vec{M}_{\cdot k}}$. For
  shift-invariant, compactly-supported PSF kernels far from the image boundary,
  the RCN is then equal to one.
\end{result}
\begin{proof}
  If there exists at least two colinear column vectors, $\vec{M}$ is \emph{rank
    deficient} and the linear inverse problem~\eqref{eq:linear-convolution} is
  ill-posed, \ie $1/\kappa(\vec{M}) = 0$. This happens when at least two sources
  occupy the same location.

  If all the sources occupy different locations, the columns are all different
  and $\vec{M}$ has maximal rank. However, a small singular value of order $k$
  means that there is $k$ column vectors that are nearly collinear. This happens
  when the corresponding sources are close enough to have their PSF affecting
  each other, \ie when their supports overlap.

  For a compactly supported PSF, disjoint supports translate in
  $\transpose{\vec{M}}\vec{M}$ being diagonal, with
  $(\transpose{\vec{M}}\vec{M})_{ii} = \Ltwo{M_{\cdot i}}^2$ being the
  eigenvalues of $\transpose{\vec{M}}\vec{M}$. The RCN of
  $\transpose{\vec{M}}\vec{M}$ is therefore the ratio between the smallest and
  the largest eigenvalues. Using
  $\kappa(\vec{M}) = \sqrt{\kappa(\vec{M}^t\vec{M})}$ (see for
  example~\cite{Gentle2017}), one concludes the result for a compactly supported
  PSF. For shift-invariant and compactly supported PSFs, the columns of
  $\vec{M}$ will have the same norm, except for sources located close to the
  image boundaries.
\end{proof}


\begin{figure}[!t]
  \centering
  \includegraphics[width=\columnwidth]{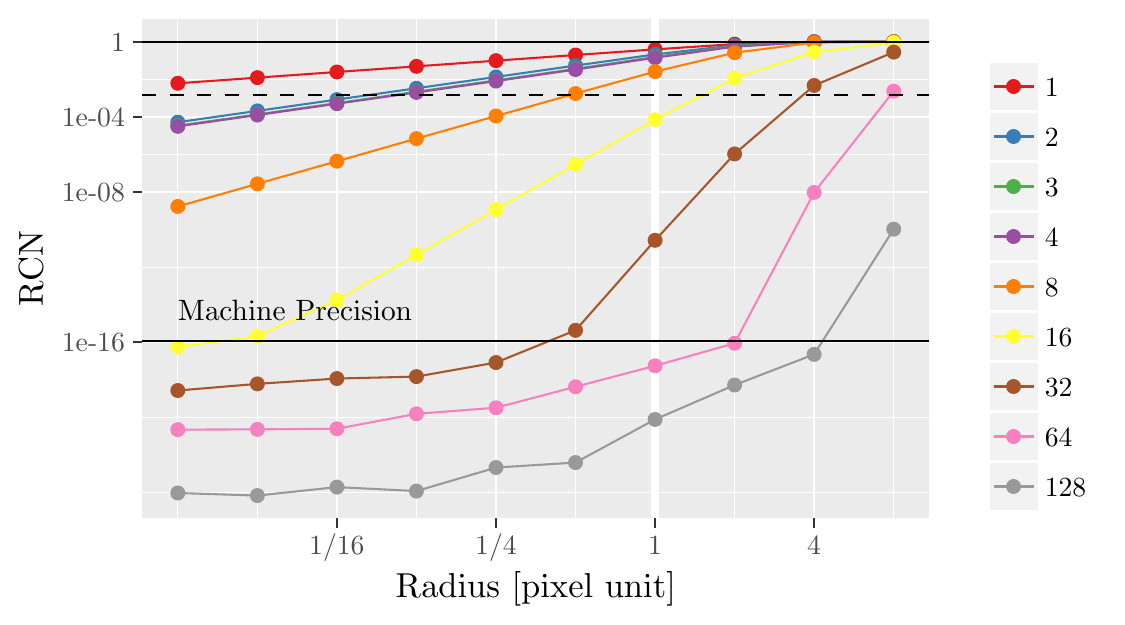}
  \caption{
    \textbf{Influence of the geometry on the reciprocal condition number}. The
    reciprocal condition number (RCN) is computed for different numbers of
    neighbors (between 1 and 128) lying on a circle of radii $2^k$ with $k$ an
    integer in $\intervalcc{-6}{3}$. One source is located at the center of the
    imaging plane, its $n$ neighbors are positioned around this central source
    according to the $n$-th roots of unity
    $(\cos 2\pi/k, \sin 2\pi/k)_{k \in 0:(n-1)}$, and scaled at different radii.
    For reference, we show the RCN of the custom convolution matrix builds on
    the whole pixel grid, with Dirichlet boundary conditions (dashed line), and
    the machine precision. The super-resolution regime starts at a radius below
    1.}
  \label{fig:kymo:rcn}
\end{figure}


%
To illustrate this theoretical result, we computed the RCN for increasing
amounts of PSFs overlap. We increase the overlap by either increasing the number
of point sources, or by reducing the distance between them. For the PSF and the
pixel size, we use the imaging parameters shown in Table~\ref{table:params}. To
make the analysis more tractable, we consider a single imaging plane. We
assemble $\vec{M}$ for different numbers and configurations of light sources. A
single source is placed at the centre of the imaging plane, and neighbouring
sources are placed on a circle according to the roots of unity. Reducing the
radius of the circle and increasing the number of sources lying on the circle
both increase the total overlap between the PSF supports.

In Fig.~\ref{fig:kymo:rcn} we observe that for the largest radius (8 pixels), a
small number of sources achieve the optimal RCN of 1. From 32 sources on, their
PSF start interacting laterally, and the amount of overlap increases with the
number of sources: the RCN decreases accordingly. When the radius decreases, the
sources interact even more. Already for two sources (\ie for one neighbouring
source, in red), the impact is moderate and the RCN decreases slowly, even at
sub-pixel distances (\ie radius below 1). However, for an increasing number of
sources, the decrease in RCN is faster with the number of sources.

\subsubsection{Resolution/well-posedness tradeoff}
Choosing a geometry to reconstruct the photometry, \ie choosing $\vec{M}$ in
equation~\eqref{eq:linear-convolution}, entails a geometric resolution defined
by the minimum distance between the point sources. However, for
diffraction-limited imaging, the PSF shape imposes a practical resolution limit,
preventing choosing arbitrarily fine geometries~\cite{Huang2010}. In
Fig.~\ref{fig:kymo:rcn} we give an inverse problem stability perspective by
showing that refining the geometric configuration brings the deconvolution
problem~\eqref{eq:linear-convolution} closer to an ill-posed problem. However,
in practice, the interest is to be able to reconstruct the intensity accurately,
especially for the kymograph problem, where the photometry reconstruction
accuracy is crucial for studying the underlying biological processes.


\begin{figure}[!t]
  \centering
  \includegraphics[width=\columnwidth]{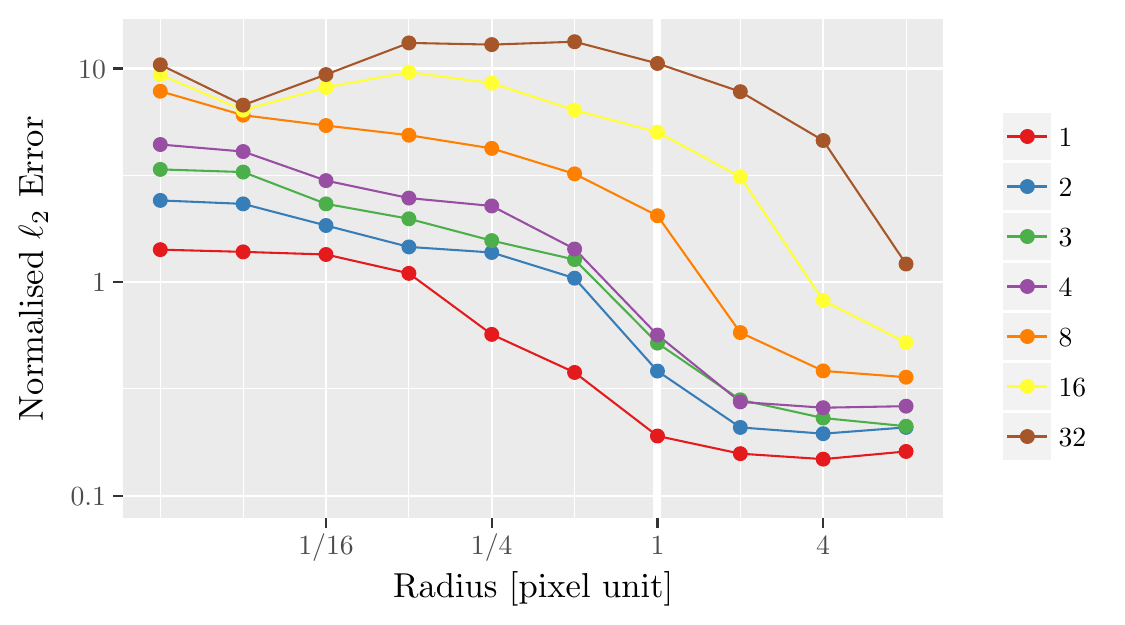}
  \caption{
    \textbf{Geometry--photometry interplay in reconstruction accuracy}. We use
    the same setup as in Fig.~\ref{fig:kymo:rcn}. The background intensity is
    uniform and set to $\mu_\textrm{bg} = 10$. Each source have an intensity of
    $\varphi_\textrm{true} = 400$. We report the median of the $\ell_2$ error
    between the estimated intensity vector $\widehat{\vec{\varphi}}$ and the
    true vector among 100 independent Poisson noise realisation. The $\ell_2$
    error is normalised by $\varphi_\textrm{true}$. Due to very bad conditioning
    beyond 32 neighbours (see Fig.~\ref{fig:kymo:rcn}) L-BFGS-B fails and the
    results are not shown.}
  \label{fig:kymo:errors}
\end{figure}


%
In Fig.~\ref{fig:kymo:errors}, we simulate for each experimental design used in
Fig.~\ref{fig:kymo:rcn} 100 independent Poisson noise realisation, where the
forward problem~\eqref{eq:linear-convolution} is used with
$\mu_\textrm{bg} = 10$, and all point sources having the same intensity,
$\varphi_\textrm{true} = 400$. For each artificial image, the maximum likelihood
estimate is computed using L-BFGS-B with positivity constraints~\cite{Byrd1995},
and the $\ell_2$ error is computed and normalised to $\varphi_\textrm{true}$. We
show the median among the 100 repeats. This analysis shows that the photometry
of sub-pixel geometric configurations is a hard inverse problem, but can still
be approached when only few sources are considered. This agrees with the
stability perspective shown in section~\ref{sec:interpl-betw-geom}. Therefore,
there is a tradeoff between the geometric resolution where one can reasonably
reconstruct photometry and well-posedness of the deconvolution inverse problem.

\subsubsection{Reconstructing kymographs sequentially is at worst ill-posed and at
  best limited by the pixel resolution}
\label{sec:reconstr-kymogr-sequ}
The previous insights allow us to motivate the challenges of the kymograph
reconstruction problem. Custom digital reconstruction algorithms amount to first
deconvolve/denoise the image on the pixel grid, and then to estimate the
photometry at any given point along the curve by reading the reconstructed
photometry at the nearest pixel. This can be formalised as a modified version of
equation~\eqref{eq:linear-convolution}:
\begin{equation}
  \label{eq:linear-convolution-with-nn}
  \def\stackalignment{c}
  \def\stacktype{L}
  \setstackgap{L}{.7\normalbaselineskip}
  \stackunder{\vec{\mu}}{\mathsmaller{n_p \times 1}}
  =
  \stackunder{\vec{\mu}_{\textrm{bg}}}{\mathsmaller{n_p \times 1}}
  +
  \stackunder{\vec{M}}{\mathsmaller{n_p \times n_p}}\,
  \stackunder{\vec{S}}{\mathsmaller{n_p \times n_s}}\,
  \stackunder{\vec{\varphi}}{\mathsmaller{n_s \times 1}}\ ,
\end{equation}
where $\vec{M}$ is now the standard digital convolution matrix, and $\vec{S}$ is
a matrix that assigns each $n_s$ position along the curvilinear geometry to its
nearest neighbour pixel. Mathematically it is a binary matrix with exactly a
single one in each column. The matrix $\vec{S}$ acts on $\vec{M}$ by selecting a
multi-set (\ie repetitions are possible) of its columns. Once this equation is
inverted for $\vec{M}$ using any standard deconvolution algorithm, $\vec{S}$ has
a trivial left inverse (its transpose) acting as a selection matrix that picks
up the nearest neighbour pixels.

However our previous analysis shows that this digital inverse
problem~\eqref{eq:linear-convolution} is either ill-posed, or the finest
kymograph resolution is limited to the pixel size. If one attempts to
reconstruct a kymograph with at least two points that corresponds to the same
nearest pixel, $\vec{S}$, and hence $\vec{MS}$ will have duplicated columns, and
therefore be ill-posed. The only way to be well-posed is to ensure to choose
points along the curve that are assigned to unique pixels. However, this has
three main limitations: the resolution cannot be finer than a pixel; the
kymograph sampling depends on the embedding of the curve in physical space; the
kymograph sampling can only be uniform for curves aligned with the image axes.

Nevertheless, usually $n_s \ll n_p$ and hence $\vec{MS}$ have much less columns
than $\vec{M}$. From the previous insights, this is a very attractive aspect.
Exploiting the knowledge of the underlying geometry of the kymograph offers the
opportunity to reduce both the number of locations needed to reconstruct the
photometry, and the amount of overlap between the associated PSF supports.
Indeed, in a 1D topology, points have less neighbours than on a 3D digital grid.
In what follows we show how to overcome these limitations and benefit from
exploiting the geometry underlying the kymograph using the virtual microscope
framework \cite{Samuylov2015}.




\section{Forward problem}
\label{sec:kymo:fwd}

\subsection{Object model}

\subsubsection{Measure-theoretic object model for incoherent imaging}
\label{sec:kymo:object-model:measure-theory}
In~\cite{Samuylov2015}, we define objects as a spatio-temporal distribution of
light sources (\emph{photometry}) restricted to a subspace of the physical space
(\emph{geometry}). In fluorescence microscopy, an object corresponds to the
spatio-temporal distribution of the fluorescently labeled proteins under
scrutiny within a biological compartment. Mathematically, this is captured by
two ingredients: for an object indexed by $\iObject$, the photometry is defined
as a positive measure in space and time encoding generalised distributions of
light sources, denoted $\photomMeas^\iObject\paren{\dy \times \dt}$, and the
geometry is encoded by a time-varying piecewise-Riemannian manifold, denoted
$\manifold^\iObject$. An object entails an \emph{object measure}, defined as the
photometry measure restricted to $\manifold^\iObject$:
\begin{equation}
  \label{eq:kymo:general-obj-measure}
  \objMeas[t, \manifold^\iObject] \paren{\dy \times \dt} \defeq
  \1{\manifold^\iObject}\paren{\y}\, \photomMeas[\t]^\iObject \paren{\dy
    \times \dt}\ ,
\end{equation}
where $\1{\manifold^\iObject}\paren{\y}$ is the indicator function assuming $1$
if $\y$ is on the manifold $\manifold^\iObject$ and $0$ otherwise. In what
follows, we assume that objects emit light continuously in time, which is
modelled as a photometry measure proportional to the Lebesgue measure $\dt$:
$\photomMeas[\t]^\iObject \paren{\dy \times \dt} = \photomMeas[\t]^\iObject
\paren{\dy} \dt$.

Fluorescence microscopy being an incoherent imaging process~\cite{Aguet2009},
the total photon flux emitted by a set of objects, denoted $\objectL$, is the
sum of the flux emitted by each individual object:
\begin{equation}
  \label{eq:kymo:superposition-obj-measure}
  \objMeas\paren{\dy\times\dt} \defeq \sum_{\mathclap{\iObject \in \objectL}} \objMeas[t, \manifold^\iObject] \paren{\dy\times\dt} \ .
\end{equation}

\subsubsection{Object model}
\label{sec:kymo:object-model}

Photons emitted by objects distant from the fluorescent marker of interest also
contribute to the total photon counts. Therefore, we model two kinds of objects:
the signal emitted by the background (due to the auto-fluorescence of the medium
and the diffuse component of the labeled proteins in the available cellular
volume) and the fluorescent proteins restricted to a curvilinear compartment.

\paragraph{Background object model}
We assume that the background signal is uniform in space but decreases in time
due to photobleaching, $\objMeas^\BG \defeq \objMeas[\t, \imageDom]^\BG$, where:
\begin{equation}
  \label{eq:kymo:object-model-bg}
  \objMeas^\BG \paren{\dy\times\dt} =
  \1{\imageDom}\paren{\y} \bgPhi \dy \dt \ ,
\end{equation}
where $\imageDom \subset \R^3$ denotes the imaging volume (subset of the
physical space), and $\bgPhi \in \Rpos$ is the background intensity. The imaging
volume $\imageDom$ is omitted because the indicator function $\1{\imageDom}$
will assume one in practice.

\paragraph{Curvilinear object model}
We assume that the curvilinear compartment is described by an open curve,
denoted \curve, and that the signal emitted by the fluorescent markers attached
along the geometry has a density with respect to the spatio-temporal Lebesgue
measure, denoted $\mtPhi\paren{\y} \in \Rpos$. We write the object measure as $\objMeas[\mathsmaller\curve] \defeq \objMeas[\t,\mathsmaller\curve]$, where:
\begin{equation}
  \label{eq:kymo:object-model-curve}
  \objMeas[\mathsmaller\curve]\paren{\dy\times\dt} =
  \1{\;\!\mathsmaller{\curve}}\paren{\y} \mtPhi \paren{\y} \dy \dt \ .
\end{equation}

\subsubsection{Kymograph--to--object mapping}
\label{sec:kymo:map:kymo-object}
We show that the notion of parameterisation of the geometry developed
in~\cite{Samuylov2015} is the key mathematical concept to define the notion of
kymograph generically.

As introduced in the virtual microscope framework~\cite{Samuylov2015}, we
parameterise the manifold encoding the geometry and use a map to embed it into
physical space. For an open curve of length $\curveLen$ at time $\t$, the
parameter space is a one-dimensional interval, denoted
$\paramSpace \defeq \intervalcc{0}{\curveLen} \subset \Rpos$. The parameter
space represents an intrinsic coordinate system attached to the curvilinear
object. The embedding is a time-varying map from the parameter space
$\paramSpace$ to the curvilinear manifold $\curve$, denoted
$\mapParamToObj: \paramSpace \to \curve \subset \imageDom$. It defines the
geometry of the object as $\curve \defeq \mapParamToObj\paren{\paramSpace}$ and
allows the intrinsic coordinate system to follow the geometry evolution in time.

In fluorescence microscopy, a kymograph represents the time evolution of the
distribution of light sources along a coordinate system intrinsic to the
curvilinear structure under scrutiny. This precisely corresponds to the time
evolution of the photometry measure in parameter space. Therefore, we call the
set of parameter spaces the \emph{kymograph space} or \emph{kymospace}, denoted
$\kymoSpace \defeq \paramSpace[\timeSet]$, where $\timeSet \subset \Rpos$ is the
time interval during which the object is observed.
The set of maps embedding the kymograph in the spatio-temporal volume, denoted
$\mapParamToObj[\mathsmaller\timeSet]$, defines a bijective mapping between a
point $\y$ on the curvilinear object $\curve$ in physical space and a point
$\paren{\t, \len}$ in kymospace, where $\len \defeq \mapParamToObjInv(\y)$.
We denote the fluorescence signal of a curvilinear object in kymospace using the
same notation as in physical space and we define it at position $(\t, \len)$ as
$ \mtPhi\paren{\len} \defeq \mtPhi\paren{\mapParamToObj\paren{\len}}$. 
Finally, we define the \emph{kymograph} on the kymospace $\kymoSpace$
as:
\begin{equation}
  \label{eq:kymo:photo-in-kymospace}
  \kymoSpace\paren*{\mtPhi[]} \defeq \set[\Big]{\paren[\Big]{\t,\,\len,\,\mtPhi\paren{\len}}{:}\ \t \in \mathcal{T},\ \len \in \paramSpace} \ .
\end{equation}

\subsubsection{Lévy process modelling the object photometry}
\label{sec:kymo:levy-process-model}

In order to develop a Bayesian formulation of the kymograph reconstruction
problem, we use the general framework of~\cite{Unser2014} to define a large
class of priors for the photometry density $\mtPhi$.

\paragraph{Reconstruction space}
We digitalise the continuous domain fluorescence signal by projecting it onto a
reconstruction space. Given a reconstruction space at resolution $\binSize$, the
approximated continuous-domain signal is~\cite{Bostan2013}:
\begin{equation}
  \label{eq:kymo:reconstruction-equation}
  \mtPhi\paren{\len} = \sum_{\iBase = 0}^{\mathclap{\nBases - 1}} \mtPhi\parenEl{\iBase}\, \basisFun\paren[\Bigg]{\frac{\len}{\binSize} - \iBase} = \transpose{\basisVec}\!\paren{\len}\, \mtPhiVec\ ,
\end{equation}
where $\basisFun$ is an interpolation basis function defined on each parameter
space,
$\nBases = \ceil{\curveLen/\binSize}$ is the number of knots,
$\cramped{\basisVec\paren{\len}\in\Rpos^{\nBases}}$ is the vector holding the
shifted/scaled basis functions,
and $\cramped{\mtPhiVec\in\Rpos^{\nBases}}$ is the vector of \emph{digital
  intensities}
$\cramped{\mtPhi\parenEl{\iBase} \defeq
  \mtPhi\paren{\len}\bigr|_{\len=\iBase\binSize}}$.
Therefore, the reconstruction of the continuous fluorescence signal along a
curvilinear object amounts to estimating a finite set of weights $\mtPhiVec$.

\paragraph{Statistical model of the photometry}
Following~\cite{Unser2014}, a large class of stochastic processes is derived
from the principle that a whitening linear operator $\whitenedOp$ transforms the
process $\mtPhi$ into a canonical \emph{Lévy innovation process} $\innov$:
\begin{equation}
  \label{eq:kymo:innov-model-continuous}
  \whitenedOp \mtPhi = \innov \ .
\end{equation}
The nature of the interpolation basis function $\basisFun$
in~\refeq{eq:kymo:reconstruction-equation} is related to the whitening operator
$\whitenedOp$ (see~\cite{Unser2014, Bostan2013}).

Following~\cite{Bostan2013}, the discretisation at resolution $\binSize$ of the
generalised stochastic process described by~\refeq{eq:kymo:innov-model-continuous}
writes:
\begin{equation}
  \label{eq:kymo:innov-model-discrete}
  \whitenedOpMat \mtPhiVec = \innovVec \ ,
\end{equation}
where $\whitenedOpMat \in \R^{\nBases \times \nBases}$ is the matrix
representation of the discrete counterpart of the operator $\whitenedOp$, and
$\innovVec \in \R^{\nBases}$ is the discrete innovation process. The statistical
features of the discretised signal and the continuous-domain innovation process
are directly related via the Lévy exponent (\cite{Bostan2013},~Theorem 3).

To simplify the inverse problem algorithm, we restrict ourselves to first-order
whitening operators. In that case, the signal $\mtPhi$ is a Lévy process, and
the interpolation basis function is a B-spline of degree zero~\cite{Bostan2013},
\ie $\basisFun(x)$ assumes $1$ if $x \in \intervalco{0}{1}$ and $0$ otherwise.
Since the basis functions are non-overlapping, the increments of the innovation
process are independent and identically distributed (see~\cite{Amini2013,
  Bostan2013}):
$\prob\paren{\vec{u}_\t} = \prod_{\iBase=0}^{\nBases - 1}
\prob_U\paren{\vec{u}_\t\parenEl{\iBase}}$.
Introducing the negative $\log$-transformed probability
$\potential \defeq -\log \prob_U$ to rewrite the latter equation in potential
form:
\begin{equation}
  \label{eq:kymo:signal-prior-dist}
  \varpi(\vec{u}_\t) \defeq - \log \prob(\vec{u}_\t) = \sum_{\iBase=0}^{\nBases - 1} \potential(\vec{u}_\t\parenEl{\iBase})\ .
\end{equation}

\subsection{Image formation model}
\label{sec:kymo:image-formation-model}
In fluorescence microscopy, the image formation process consists of two steps.
First, the signal emitted by fluorescently labeled objects is distorted due to
the random nature of light emission, the propagation through the environment
(\eg cytoplasm, coverslip, immersion layer, optics in the microscope objective),
and the sampling at the pixel grid. Then, the signal is corrupted by the
measurement noise (\eg spurious charge, amplification noise, readout noise) and
the quantisation at the camera detector incurred by its encoding into grey
values.

\subsubsection{Object--to--pixel mapping}
\label{sec:kymo:map:object-pixel}

\paragraph{General object--to--pixel mapping}
We assume that the microscope is operated in a regime where Poisson shot noise
(due to the random nature of photon emission) dominates the photon counting
statistics~\cite{Snyder1993}. Therefore, we assume that the number of photons
collected at pixel $\px_\iPx \subset \imageDom$ at time
$\t_\mathsmaller{\!\!\:\iFrame}$ during \tA (\ie during the interval
$\timeSet_{\discTime, \tA} \defeq \tAin[\discTime]$) is a random variable
following a Poisson distribution parameterised by the \emph{total expected
  photon count}
$\expPhCount$, \ie $\obsPhCount \sim \Poisson\paren{\expPhCount}$.
We also assume that the family of random variables
$\set{\obsPhCount}_{\iFrame\iPx}$ is mutually independent, but not identically
distributed. Indeed, the optical distortions correlate the photon statistics in
space. This is captured by PSF kernel, denoted $\psfFun$. We assume that the PSF
is shift-invariant, hence acting as a convolution operator on the object
measure. The resulting measure characterises the expected photon flux in space
and time:
$\phFlux\paren{\objMeas} \defeq \psfFun \conv \objMeas$, 
where the convolution is understood in the sense of measures,
(see~\cite{Samuylov2015}). The expected number of photons collected on the pixel
surface $\px_\iPx$ during the acquisition interval $\timeSet_{\discTime, \tA}$
corresponds to the integral of the convolution measure:
\begin{equation*}
  \expPhCount\paren{\objMeas} \defeq
  \int_{\px_\iPx \times \timeSet_{\discTime, \tA}} \paren{\psfFun \conv
    \objMeas}\paren{\dx\times\dt}\ .
\end{equation*}
The latter operator is linear in the object measure, thus the superposition
principle for the object measures~\refeq{eq:kymo:superposition-obj-measure}
entails the linearity of the expected number of photons:
\begin{equation}
  \label{eq:kymo:linearity-expected-photon-count}
  \expPhCount\paren*{\sum_{\iObject\in\objectL} \objMeas[\t,\manifold^\iObject]} = \sum_{\iObject\in\objectL} \expPhCount\paren{\objMeas[\t,\manifold^\iObject]}\ .
\end{equation}
Therefore, to compute the total expected photon count, we can consider each
object separately.
In the following, we denote the contribution to the expected photon flux coming
from the $\iObject$-th object by
$\phFlux[\iObject]_\t \defeq \phFlux\paren{\objMeas[\!\manifold^\iObject]}$.

\paragraph{Background and curvilinear objects}
Following~\cite{Snyder1993}, we normalise the PSF kernel to a
probability. Therefore, the contribution from the background object described
by~\refeq{eq:kymo:object-model-bg} to the total expected photon flux writes as:
\begin{equation}
  \label{eq:kymo:obj-bg}
  \phFlux[\BG]_\t\paren{\dx\times\dt} = \bgPhi \dx \dt\ \eqdef \phFlux[\BG](\t)\dx\dt\ .
\end{equation}

For the curvilinear object~\refeq{eq:kymo:object-model-curve}, the photon flux
measure has a density with respect to the spatio-temporal Lebesgue measure:
$\phFlux[\C]_\t\paren{\dx\times\dt} \defeq \phFlux[\C]\paren{\x,\t}\dx\dt$. This
density involves an integral on the curve \curve:
\begin{equation*}
  \phFlux[\C]\paren{\x,\t} = \int_{\curve} \mtPhi\paren{\y} \psfFun\paren{\x - \y} \dy\ .
\end{equation*}
The latter formula requires potentially integrating over the three dimensional
imaging volume. However, using the bijective parameterisation introduced
in~\refsection{sec:kymo:map:kymo-object}, we pull back the integration into the
parameter space, thus reducing the dimensionality of the integral:
\begin{equation}
  \label{eq:kymo:ph-flux-c}
  \phFlux[\C](\x,\t) = \int_{\paramSpace} \mtPhi\paren{\len} \psfFun\paren{\x - \mapParamToObj\paren{\len}}\, \RiemannianMetric\paren{\len} \dlen\ ,
\end{equation}
where $\RiemannianMetric$ is the Riemannian metric induced by $\mapParamToObj$
and defined as the Euclidean norm of the derivative along the curve of the
parameterisation:
$\RiemannianMetric \defeq \abs*{\diff/\dlen\ \mapParamToObj}_2$.
The total expected photon flux sampled by the pixel array is a spatio-temporal
measure with density:
$\phFlux(\x,\t) \defeq \phFlux[\BG](\t)+\phFlux[\C](\x,\t)$.

\paragraph{Discretised object--to--pixel mapping}
In order to discretise the expected photon flux, two levels of integration need
to be approximated: the object-level integration and the sampling integration on
the pixel array. The former amounts to the virtual source approximation
introduced in~\cite{Samuylov2015}. It involves approximating an integral in
parameter space and the integration during the acquisition interval
(see~\cite{Samuylov2015}). The convolution integral for a curvilinear objects
requires a one dimensional quadrature:
\begin{equation}
  \label{eq:kymo:obj-mt}
  \phFlux(\x,\t) \approx \bgPhi +
  \sum_{\iVS \in \VS_\t} \weightVS[\t, \iVS]\paren{\len_\iVS}\, \mtPhi\paren{\len_\iVS}\,
  \psfFun\paren{\x - \mapParamToObj\paren{\len_\iVS}} \ ,
\end{equation}
where $\VS_{\t}$ is the set of $\nVS$ virtual point source indices, and the weight
function is the product of the Riemannian metric accounting for the geometry and
the quadrature weight, denoted
$\weightQ[\t, \iVS]$: $\weightVS[\t, \iVS]\paren{\len_\iVS} \defeq  \RiemannianMetric\paren{\len_\iVS}\,\weightQ[\t, \iVS]$.
The \emph{virtual source approximation} amounts to approximating the integral of
the expected photon flux $\phFlux$ during the acquisition time interval. We
choose a \emph{right continuous with left limits} piecewise constant
approximation in time for the quadrature. This means that any function in time
is approximated by its value at the beginning of each integration interval:
\begin{equation*}
  \phFlux(\dx \times \timeSet_{\discTime, \tA}) = \int_{\discTime}^{\discTime+\tA}\!\!\! \phFlux(\x,t)\dt \ \dx \approx \tA\, \phFlux(\x, \discTime)\dx\ .
\end{equation*}

To approximate the expected photon count at pixel $\px_\iPx$, we use a midpoint
rule to integrate the previous equation:
\begin{equation}
  \label{eq:kymo:ifm-general-integrated}
  \expPhCount \approx \pxArea \tA\, \phFlux\paren{\discPos,\discTime} \eqdef \phFluxInt\paren{\discPos, \discTime}\ ,
\end{equation}
where $\discPos$ is the centre of \iPx-th pixel, and the bar over a quantity $q$
denotes integration in space and time, \ie
$\integrated{q} \defeq \pxArea \tA\, q$. Using~\refeq{eq:kymo:obj-mt}, we obtain the
following approximation of the expected photon flux:
\begin{equation*}
  \phFluxInt\paren{\discPos,\discTime} = \bgPhiInt[\iFrame] +
  \sum_{\iVS \in \VS_\iFrame} \weightVSInt[\iFrame, \iVS]\paren{\len_\iVS}\, \mtPhi[\iFrame]\paren{\len_\iVS}\,
  \psfFun\paren{\discPos - \mapParamToObj[\iFrame]\paren{\len_\iVS}} \ .
\end{equation*}

To obtain the final \emph{digital approximation} of the expected photon count,
denoted $\expPhCountVec \in \Rpos^\nPx$, we
stack~\refeq{eq:kymo:ifm-general-integrated} into a vector storing the expected
photon count of the \nPx pixels at time $\discTime$:
\begin{equation*}
  \expPhCountVec\paren{\mtPhi[\iFrame]} = \bgPhiInt[\iFrame]\,\onesVec_\nPx +
  \sum_{\iVS \in \VS_\iFrame} \weightVSInt[\iFrame,\iVS]\paren{\len_\iVS}\, \mtPhi[\iFrame]\paren{\len_\iVS}\,
  \psfVec[\iFrame]\paren{\len_\iVS} \ ,
\end{equation*}
where $\onesVec_\nPx \in \set{1}^\nPx$ is the vector of ones,
$\psfVec[\iFrame]\paren{\len_\iVS} \in \Rpos^\nPx$ is the vector with $\iPx$-th
element defined as
$\psfFun\paren{\discPos - \mapParamToObj[\iFrame]\paren{\len_\iVS}}$.
Inserting~\refeq{eq:kymo:reconstruction-equation} and reordering the terms, we
obtain:
\begin{equation*}
  \expPhCountVec\paren{\mtPhiVec} = \bgPhiInt[\iFrame]\,\onesVec_\nPx +
  \sum_{\iVS \in \VS_\iFrame} \psfVec[\iFrame]\paren{\len_\iVS}\, \weightVSInt[\iFrame,\iVS]\paren{\len_\iVS}\, \transpose{\basisVec[\iFrame]}\!\paren{\len_\iVS}\, \mtPhiVec[\iFrame]\ .
\end{equation*}
Introducing the convolution matrix:
\begin{equation*}
  \psfMat[\iFrame] \defeq \left[\psfVec[\iFrame]\paren{\len_1} \cdots \psfVec[\iFrame]\paren{\len_\iVS} \cdots \psfVec[\iFrame]\paren{\len_{\nVS[\iFrame]}} \right] \in\Rpos^{\nPx\times\nVS[\iFrame]}\ ,
\end{equation*}
the basis matrix:
\begin{equation*}
  \basisMat[\iFrame] \defeq \left[\basisVec[\iFrame]\paren{\len_1} \cdots \basisVec[\iFrame]\paren{\len_\iVS} \cdots \basisVec[\iFrame]\paren{\len_{\nVS[\iFrame]}} \right]  \in\Rpos^{\nBases\times\nVS[\iFrame]}\ ,
\end{equation*}
and stacking the integrated weights into a vector, denoted $\weightVSVecInt$, we
finally write the digital expected photon count vector at time $\discTime$ as:
\begin{equation}
  \label{eq:kymo:expected-photon-count-vector}
  \def\stackalignment{c}
  \setstackgap{L}{12pt}
  \def\stacktype{L}
  \expPhCountVec\paren{\mtPhiVec[\iFrame]} = \bgPhiInt[\iFrame]\,\onesVec_\nPx +
  \stackunder{\psfMat[\iFrame]}{\mathsmaller{\nPx \times \nVS[\iFrame]}}
  \stackunder{\diag\paren{\weightVSVecInt[\iFrame]}}{\mathsmaller{\nVS[\iFrame] \times \nVS[\iFrame]}}
  \stackunder{\transpose{\basisMat[\iFrame]}}{\mathsmaller{\nVS[\iFrame] \times \nBases[\iFrame]}\hspace{0.2cm}}
  \stackunder{\mtPhiVec[\iFrame]}{\mathsmaller{\nBases[\iFrame] \times 1}}\ .
\end{equation}
The linear part of the latter equation represents a mapping between the
intensity vector along the curvilinear object in kymospace and the expected
photon count vector parameterising the Poissonian photon statistics:
$ \kymoToImageMat[\iFrame] \defeq \psfMat[\iFrame]
\diag\paren{\weightVSVecInt[\iFrame]} \transpose{\basisMat[\iFrame]}$.

\subsubsection{Pixel--to--image mapping}
\label{sec:kymo:map:pixel-image}
The pixel--to--image mapping models the conversion of photons hitting the camera
into grey values. This mapping, denoted $\mapPxToImg$, is camera-specific. In
this work, we assume an electron multiplication charge-coupled device (EM CCD).
For the sake of simplicity, we model this mapping deterministically:
\begin{equation}
  \label{eq:kymo:pixel-to-image-map}
  \obsGVCount  = \quantEfficiency\, \mGain\, \ADU^{-1} \obsPhCount\, + \cameraBias \eqdef \nu\paren{\obsPhCount}\ ,
\end{equation}
where $\quantEfficiency$ is the quantum efficiency, $\mGain$ is the
multiplication gain, $\ADU$ is the analog-to-digital proportionality factor, and
$\cameraBias$ is the camera bias.

\section{Inverse problem}
\label{sec:kymo:inv}

In this work, we focus on estimating the distribution of light sources in
kymospace. Therefore, we assume that the dynamics of the geometry of a
curvilinear object and of the background intensity are estimated beforehand. We
formulate the inverse problem to reconstruct the kymograph along a curvilinear
object from a sequence of $\nFrames$ images. At each time point
$\t_\iFrame \defeq (\iFrame-1)\,\tA$ for
$\iFrame \in \frameSet \defeq \set{1,\dots, \nFrames}$, the signal is collected
at $\nSlices$~focal planes resulting in a stack of images, denoted
$\obsGVMat[\iFrame] \in \Z^{\nSlices\times\nHeight\times\nWidth}$, and
$\nHeight \times \nWidth$ is the number of pixels in camera array.
We make the standard assumption of neglecting the objects dynamics during the
acquisition of a stack, thus the number of pixels in
\refeq{eq:kymo:expected-photon-count-vector} is
$\nPx \defeq \nSlices\, \nHeight\, \nWidth$.
The kymograph space and its embedding are defined by the estimated geometry of
the curvilinear object:
$\kymoSpace_\frameSet = \set{\paramSpaceHat[\iFrame]}_{\iFrame \in \frameSet}$
and
$\mapParamToObj[\frameSet] = \set{\mapParamToObjHat[\iFrame]}_{\iFrame \in
  \frameSet}$, respectively.

\subsection{Nearest neighbour (NN) kymograph}
\label{sec:kymo:NN}

A straightforward and widespread kymograph estimate is to sample the grey values
along the estimated curvilinear geometry. Each point
$\paren{\t_\iFrame,\, \len_\iVS}$ in kymospace is embedded in physical space by
$\mapParamToObjHat[\!\iFrame]$, and then the nearest pixel centre index is
selected:
$\iPx_{\iFrame\iVS} \defeq \arg\min_{\iPx \in \iPxSet}
\Ltwo*{\mapParamToObjHat[\!\iFrame]\paren{\len_{\iVS}} - \discPos}$,
where $\iPxSet \defeq \set{1, ..., \nPx}$ is the set of all pixels in the image
stack.
The \emph{nearest neighbour kymograph} assigns the signal intensity to the grey
value of the selected pixel:
\begin{equation*}
  \kymoNN \defeq \set*{\paren*{\t_\iFrame,\ \len_\iVS, \, \obsGVMat[\iFrame]\parenEl{ \iPx_{\iFrame\iVS}} }:\ \t_\iFrame \in \frameSet,\ \len_\iVS \in \paramSpaceHat[\iFrame]}\ .
\end{equation*}
The nearest neighbour estimate suffers from all the distortions and degradation
affecting the image space, \eg sampling resolution limit, blurring,
photobleaching, measurement noise.

\subsection{Sub-pixel resolution kymograph: a MAP formulation}
\label{sec:kymo:MAP}

We aim at reconstructing the photometry dynamics along an estimated curvilinear
geometry:
\begin{equation*}
  \kymoMAP \defeq \set*{\paren*{\t_\iFrame,\ \len_\iVS,\, \mtPhiMAP[\iFrame,\iVS]}:\ \iFrame \in \frameSet,\ \len_\iVS \in \paramSpaceHat[\iFrame]}\ ,
\end{equation*}
where from~\refeq{eq:kymo:reconstruction-equation}:
$\mtPhiMAP[\iFrame,\iVS] \defeq
\transpose{\basisVec[\iFrame]}\paren{\len_\iVS}\, \mtPhiVecMAP[\iFrame]$.
The vector of digital intensities
$\mtPhiVecMAP[\iFrame] \in \Rpos^{\nBases[\iFrame]}$ is estimated as the
solution of the MAP problem associated to the forward problem described in
\refsection{sec:kymo:fwd}.

Following~\cite{Bostan2013}, the MAP associated to the Lévy innovation process
$U$ and the whitening operator $\whitenedOpMat[\iFrame] \in \R^{\nBases[\iFrame]
\times \nBases[\iFrame]}$ writes as the following minimisation problem:
\begin{equation}
  \label{eq:kymo:MAP}
  \mtPhiVecMAP[\iFrame] = \arg\min_{\mtPhiVec[] \in \Rpos^{\nBases[\iFrame]}} \nll\paren{\mtPhiVec[] \, | \, \obsGVMat[\iFrame]} + \potential \paren*{ \whitenedOpMat[\iFrame] \mtPhiVec[]} \ ,
\end{equation}
where
$\nll\paren{\mtPhiVec[] | \obsGVMat[\iFrame]} = - \log \prob
\paren{\obsGVMat[\iFrame] | \mtPhiVec[]}$ is the negative log-likelihood
function, and $\potential$ is the potential function encoding the prior about
the fluorescence signal described in \refeq{eq:kymo:signal-prior-dist}.

The negative log-likelihood derives from the Poissonian assumption and the
linear and deterministic pixel--to--image mapping (see
\refsection{sec:kymo:map:object-pixel} and \refsection{sec:kymo:map:pixel-image}).
Indeed, for a given image $\obsGVMat[\iFrame]$, the invertible pixel--to--image
mapping $\mapPxToImg$, allows converting grey values into photon counts that we
stack into a vector
$\obsPhCountVec[\iFrame] \defeq
\vectorise\paren*{\mapImgToPx\paren{\obsGVMat[\iFrame]}}$. 
The negative log-likelihood writes as
$\nll\paren{\mtPhiVec[] \,|\, \obsGVMat[\iFrame]} = \nll\paren{\mtPhiVec[] \,|\,
  \obsPhCountVec[\iFrame]}$, where:
\begin{equation}
  \begin{aligned}
    \nll\paren{\mtPhiVec[] \,|\, \obsPhCountVec[\iFrame]}
    &= \inner*{\onesVec_\nPx,\, \obsPhCountVec[\iFrame] \log{ \frac{\obsPhCountVec[\iFrame]}{\expPhCountVec[\iFrame]\paren{\mtPhiVec[]}} } + \expPhCountVec[\iFrame]\paren{\mtPhiVec[]} - \obsPhCountVec[\iFrame] }\\
    &\eqdef \nPx\,
    \nllNormed\paren*{\bgPhiInt[\iFrame]\,\onesVec_\nPx+\kymoToImageMat[\iFrame]\mtPhiVec[]}\ .
  \end{aligned}
\end{equation}
In the latter equation we define the normalised likelihood $\nllNormed$, where
the number of pixels $\nPx$ is used as a normalisation to make $\nll$ values
comparable between different data. Similarly, we normalise the potential
function to account for a varying object length, \ie
$\potential \eqdef \nBases[\iFrame]\, \potentialNormed[\iFrame]$.

To constrain the fluorescence signal to positive values, we introduce the
indicator function of the set
$\inidicatorSet_\iFrame \defeq \Rpos^{\nBases[\iFrame]}$, denoted
$\indicatorOptim_{\mathcal{S}_\iFrame}\paren{\mtPhiVec[]}$, as the function
assuming $0$ if $\mtPhiVec[]$ belongs to $\inidicatorSet_\iFrame$, and $+\infty$
otherwise. The final MAP optimisation problem writes:
\begin{equation}
  \label{eq:kymo:optim-final}
  \begin{aligned}
    \mtPhiVecMAP[\iFrame]
    \defeq & \arg\min_{\mtPhiVec[]}\,
    \nllNormed\paren*{\bgPhiInt[\iFrame]\,\onesVec_\nPx+\kymoToImageMat[\iFrame]\mtPhiVec[]} + \\
    & \qquad \qquad \qquad \qquad
    \potentialNormed[\iFrame]\paren*{\whitenedOpMat[\iFrame] \mtPhiVec[]}  +
    \indicatorOptim_{\inidicatorSet_\iFrame}\paren{\mtPhiVec[]} \\
    \eqdef & \arg\min_{\mtPhiVec[]}\,
    \energyMAP_\iFrame\paren*{\bgPhiInt[\iFrame]\,\onesVec_\BG + \operator_\iFrame\mtPhiVec[]}\ ,
  \end{aligned}
\end{equation}
where $\onesVec_\BG$ is the vector selecting the background components,
implemented by stacking the vector of ones $\onesVec_\nPx$ with the vector of
zeros $\zerosVec_{2\nBases[\iFrame]}$. The latter equation highlights the nature
of this optimisation problem as a sum of three functionals in $\mtPhiVec[]$
coupled through their arguments by the following operator:
$\operator_\iFrame \defeq \transpose{
  \begin{bmatrix}
    \transpose{\kymoToImageMat[\iFrame]} 
    \transpose{\whitenedOpMat[\iFrame]} 
    \eyeMat_\iFrame                 
  \end{bmatrix}}$,
where $\eyeMat_\iFrame \in \R^{\nBases[\iFrame]\times\nBases[\iFrame]}$ is the
identity matrix. The convexity of problem~\refeq{eq:kymo:optim-final} depends
only the convexity of the innovation $\potentialNormed[\iFrame]$ (see Table I
in~\cite{Bostan2013}).

\subsection{Fully-split formulation of the algorithm}

\begin{figure}[!t]
  \removelatexerror
  \begin{algorithm}[H]
    \label{alg:full}
    \caption{Fully-split ASB for solving \eqref{eq:kymo:optim-final}}
    \SetKwInOut{Input}{Input}
    \SetKwInOut{Output}{Output}

    \Input{ $\obsGVMat[\iFrame]$, $\curve[\iFrame]$, $\bgPhiInt[\iFrame]$, $\step$}
    \BlankLine
    \Output{$\mtPhiVecMAP[\iFrame] \defeq \mtPhiP^\infty$}
    \BlankLine
    $ \w^0_{1} = \obsPhCountVec, \
    \w^0_{2} =
    \w^0_{3} = \zerosVec_{\nBases[\iFrame]}, \
    \b^0 =\zerosVec_{\nPx+2\nBases[\iFrame]}$
    \BlankLine

    \While{\textsc{not converged}}{
      Least squares sub-problem:

      \begin{fleqn}[\dimexpr(\leftmargini-\labelsep)/8]
        \begin{gather}
          \label{eq:kymo:sp-ls}
          \mtPhiLSn = \arg\min_{\mtPhiLS} \Ltwo*{\bc + \bgPhiInt[\iFrame]\, \onesVec_\BG + \operator_\iFrame \mtPhiLS - \wc}^2
        \end{gather}
      \end{fleqn}

      $\nllNormed$ sub-problem:

      \begin{fleqn}[\dimexpr(\leftmargini-\labelsep)/8]
        \begin{gather}
          \label{eq:kymo:sp-nll}
          \mtPhiNLLn =  \Prox_{\step\, \nllNormed} \paren*{\bMtPhiNLLc + \expPhCountVec\paren{\mtPhiVec[]^{\iIter+1}}}
        \end{gather}
      \end{fleqn}

      Innovation potential ($\potentialNormed[\iFrame]$) sub-problem:

      \begin{fleqn}[\dimexpr(\leftmargini-\labelsep)/8]
        \begin{gather}
          \label{eq:kymo:sp-innov}
          \mtPhiTn = \Prox_{\step\, \potentialNormed[\iFrame]} \paren*{\bMtPhiTc + \whitenedOpMat[\iFrame] \mtPhiLSn}
        \end{gather}
      \end{fleqn}

      Positivity constraint sub-problem:

      \begin{fleqn}[\dimexpr(\leftmargini-\labelsep)/8]
        \begin{gather}
          \label{eq:kymo:sp-proj}
          \mtPhiPn = \Proj_{\mathcal{S}_\iFrame} \paren*{\bMtPhiPc + \mtPhiLSn}
        \end{gather}
      \end{fleqn}

      Bregman update (dual gradient ascent):

      \begin{fleqn}[\dimexpr(\leftmargini-\labelsep)/8]
        \begin{equation}
          \label{eq:kymo:breg-update}
          \bn = \bc + \bgPhiInt[\iFrame]\, \onesVec_\BG + \operator_\iFrame \mtPhiLS^{\iIter+1} - \wn
        \end{equation}
      \end{fleqn}
    }
  \end{algorithm}
\end{figure}


In order to exploit the additive structure of the optimisation
problem~\refeq{eq:kymo:optim-final}, we use an operator splitting strategy based
on the alternating split Bregman (ASB) algorithm (\eg see ~\cite{Goldstein2009}
when $\potential$ is the $\ell_1$ norm). We use a fully-decoupled strategy as
introduced in~\cite{Setzer2010} and advocated in~\cite{Paul2013}. The strategy
involves three steps. The first step is to write~\refeq{eq:kymo:optim-final} as
a sum of two functionals:
\begin{equation*}
    \mtPhiVecMAP[\iFrame] = \arg\min_{\mtPhiVec[]}\,
    \inner*{\zerosVec_{\nBases[\iFrame]}, \mtPhiVec[]} +
    \energyMAP_\iFrame\paren{
    \bgPhiInt[\iFrame]\,\onesVec_\BG + \operator_\iFrame \mtPhiVec[]
    }\ .
\end{equation*}
The second step introduces a set of decoupling variables:
\begin{equation}
  \label{eq:kymo:dummy-var}
  \vec{w} = \transpose{
  \begin{bmatrix}
    \transpose{\mtPhiNLL}
    \transpose{\mtPhiT}
    \transpose{\mtPhiP}
  \end{bmatrix}}
  \defeq \bgPhiInt[\iFrame]\,\onesVec_\BG + \operator_\iFrame \mtPhiVec[]\ ,
\end{equation}
and a Bregman proximal point algorithm to enforce the constraint:
\begin{equation*}
  \begin{aligned}
    \paren*{\wn, \mtPhiVec[]^{\iIter+1}}
    & = \arg\min_{\w, \mtPhiVec[]}\ \inner*{\zerosVec_{\nBases[\iFrame]},
    \mtPhiVec[]} + {}\\
    & \energyMAP_\iFrame\paren{\w} +
    \frac{1}{2\step} \Ltwo*{\bc + \bgPhiInt[\iFrame]\,\onesVec_\BG +  \operator_\iFrame \mtPhiVec[] - \w}^2\\
    \bn & = \bc + \bgPhiInt[\iFrame]\, \onesVec_\BG + \operator_\iFrame
    \mtPhiLS^{\iIter+1} - \wn\ .
  \end{aligned}
\end{equation*}
The third step amounts to solving the latter optimisation problem with an
alternating minimisation strategy, resulting in~\refalg{alg:full}.

To ensure that \refalg{alg:full} yields a positive estimate for the fluorescence
intensity $\mtPhiVecMAP[\iFrame]$, we output the result of the projection step
at convergence, denoted $\mtPhiP^\infty$. This strategy results in a
least-squares problem, a proximal evaluation problem, and a Bregman linear
update. The proximal operator of a function $g$ is defined as:
$\Prox_{g}(x) \defeq \arg\,\min_y\, g(y)+\frac{1}{2} \Ltwo{x-y}^2$. The proximal
of $\iota_{\mathcal{S}}$ is the projection onto this set:
$\Prox_{\iota_\mathcal{S}} = \Proj_{\mathcal{S}}$.

\subsubsection{Least squares sub-problem \refeq{eq:kymo:sp-ls}}

The least-squares sub-problem amounts to solving the following normal equations:
\begin{equation*}
  \transpose{\operator_\iFrame} \operator_\iFrame\, \mtPhiLSn =
  \transpose{\operator_\iFrame}\paren*{\wc - \paren*{\bc + \bgPhiInt[\iFrame]\, \onesVec_\BG}}\ .
\end{equation*}
In digital image processing, the geometry of the pixel grid allows solving
efficiently the normal equations using a spectral method (\eg applying discrete
cosine transform for Neumann boundary conditions, see \eg \cite{Setzer2010}).
However, the virtual sources are not aligned with the pixel grid, preventing the
use of spectral methods. Nevertheless, following~\cite{Samuylov2015}, it is
possible to efficiently compute the operator $\kymoToImageMat[\iFrame]$ using
the improved fast Gaussian transform (IFGT, see~\cite{Morariu2008}). For
computational efficiency, the matrix inverse:
\begin{equation*}
  \inverse{\transpose{\operator_\iFrame} \operator_\iFrame} = \inverse{
  \transpose{\kymoToImageMat[\iFrame]} \kymoToImageMat[\iFrame] +
  \transpose{\whitenedOpMat[\iFrame]} \whitenedOpMat[\iFrame] +
  \eyeMat_\iFrame
  } \in \R^{\nBases[\iFrame]\times\nBases[\iFrame]} \ ,
\end{equation*}
is pre-computed outside the main iteration loop.

\subsubsection{Likelihood sub-problem \refeq{eq:kymo:sp-nll}}

The solution of the second sub-problem requires solving the following quadratic
equation for each pixel, independently (see \eg \cite{Setzer2010,Paul2013}):
\begin{equation*}
  \paren*{w_{1\iPx}^{\iIter+1}}^2 +
  w_{1\iPx}^{\iIter+1} \paren*{\frac{\step}{\nPx} - \paren*{b_{1\iPx}^\iIter +
  \expPhCount[\iFrame\iPx]\paren{\mtPhiLSn}}} - \frac{\step}{\nPx} n_{\iFrame\iPx} = 0 \ .
\end{equation*}
This quadratic equation has two roots (positive discriminant) and always admits
a positive one because the product of its roots is negative (\ie
$- \frac{\step}{\nPx} n_{\iFrame\iPx}<0$). The admissible solution
$w_{1\iPx}^{i+1}$ is the positive root because this variable is related to the
expected photon count \refeq{eq:kymo:expected-photon-count-vector} via the
constraint \refeq{eq:kymo:dummy-var}.

\subsubsection{Innovation potential sub-problem \refeq{eq:kymo:sp-innov}}
The solution depends on the innovation process prior. For example, if the
innovation process is driven by the Laplace distribution,
$\potential\paren{\mtPhiT} = \reg \Lone{\mtPhiT}$, then the proximal operator is
a component-wise soft thresholding:
\begin{equation*}
  w_{2\iBase}^{\iIter+1} = \shrink\paren*{b_{2\iBase}^\iIter + (\whitenedOpMat[\iFrame] \mtPhiLSn)\parenEl{\iBase}, \frac{\step\reg}{\nBases[\iFrame]}} \ ,
\end{equation*}
where $\shrink(w, c) \defeq \max(\abs{w}-c,0)\sign(w)$. If the innovation
process is driven by the Gaussian distribution,
$\potential\paren{\mtPhiT} = \reg \Ltwo{\mtPhiT}$, then the shrinkage operator
is given by $\shrink(w, c) \defeq w/\paren{1 + 2c}$, see~\cite{Parikh2014}. Both
potentials introduce a regularisation parameter, denoted $\reg$, that controls
the tradeoff between the data fidelity term $\nllNormed$ and the belief in the
prior about the underlying innovation process $\potentialNormed[\iFrame]$.

\subsubsection{Positivity constrain sub-problem \refeq{eq:kymo:sp-proj}}
This sub-problem decouples and is computed as the projection onto $\Rpos$:
\begin{equation*}
  \mtPhiPn = \max{\paren{\zerosVec, \, \bMtPhiPc + \mtPhiLSn}} \ ,
\end{equation*}
where $\max$ is understood component-wise.




\section{Experiments}
\label{sec:kymo:exp}

In this section, we apply our framework to reconstruct the spatio-temporal
distribution of light sources along curvilinear biological structures such as
microtubules (MTs).
We start with discussing the difficulties associated to estimating kymographs in
a challenging biological problem: studying microtubule dynamics \invivo.
Then, we use the virtual microscope framework introduced in~\cite{Samuylov2015}
to compute synthetic image data for different geometry and photometry scenarios.
We use these data to demonstrate the capabilities of our MAP framework and
compare it with the solution of two other inverse problems: the nearest
neighbour signal reconstruction and the fitting of a known parametric signal
model.
Finally, we apply our framework to a real-world data set consisting of
time-lapse images of pre-anaphase microtubule dynamics in a model system: the
budding yeast \SCerevisiae. We show that our framework allows reconstructing
details that are not accessible to conventional digital techniques. Given the
estimated sub-pixel resolution kymograph as ground truth objects, we use the
virtual microscope framework to simulate the real data and investigate the
robustness of our framework to noise, sampling and regularisation.

\subsection{MT dynamics \invivo: the need for better kymographs}
\label{sec:kymo:MT}

Microtubules are highly dynamic polar filaments that are crucial for the cell
viability~\cite{Alberts2014}. Their main feature is a stochastic transition
between growth and shrinkage that results in tightly regulated outcomes:
microtubules are essential for the intra-cellular organisation, transport, and
cell division requiring microtubules to be correctly positioned in space and
time. This tight regulation is achieved by numerous microtubule-associated
proteins, including plus-end-tracking proteins that accumulate at the
microtubule end exposed to the cytoplasm~\cite{Akhmanova2015}. However, many
regulatory mechanisms of microtubule dynamics remain largely unknown, making
this topic a fundamental problem in cell biology~\cite{Mennella2005,Borisy2016}.

Kymographs are considered to be a \emph{de facto} standard data representation
to study microtubule dynamics \invivo and \invitro. They are used to infer
parameters quantifying the dynamics at the microtubule ends
\cite{Mennella2005,Kner2009,Smal2010,Mangeol2016}. Moreover, they are used to
reconstruct trajectories of motor proteins, cargos and vesicles moving along the
microtubule lattice~\cite{Racine2007,Chenouard2010,Zhang2011}, estimate
parameters quantifying their dynamics
\cite{Mangeol2016,Chaphalkar2016,Neumann2017}, and study the underlying
regulatory mechanisms~\cite{Racine2007,Bieling2007}. Indeed, the analysis of
kymographs provides a wealth of quantitative information such as the
orientation, the velocity, and the pausing times of the transported
particles~\cite{Welzel2009}, the co-localization of motor proteins and
microtubule-associated proteins~\cite{Bieling2007,Roberts2014}.

This paradigm relies on the hypothesis that the signal displayed in the
kymograph reflects the spatio-temporal dynamics of the fluorescently-labeled
structures. However, as we described in \refsection{sec:kymo:fwd}, many
distortions impair the relationship between the labeled objects and the image
data.

Our Bayesian framework accounts for the distortions of the fluorescence signal
explicitly in the modelling of the forward problem, and uses a general class of
signal priors to cope with these imaging limitations. We start by assessing the
capability of our framework on synthetic data generated by the virtual
microscope framework.

\subsection{Reconstructing light source distributions for different geometry and
photometry scenarios}


\begin{figure*}[!t]
  \centering
  \includegraphics[width=\textwidth]{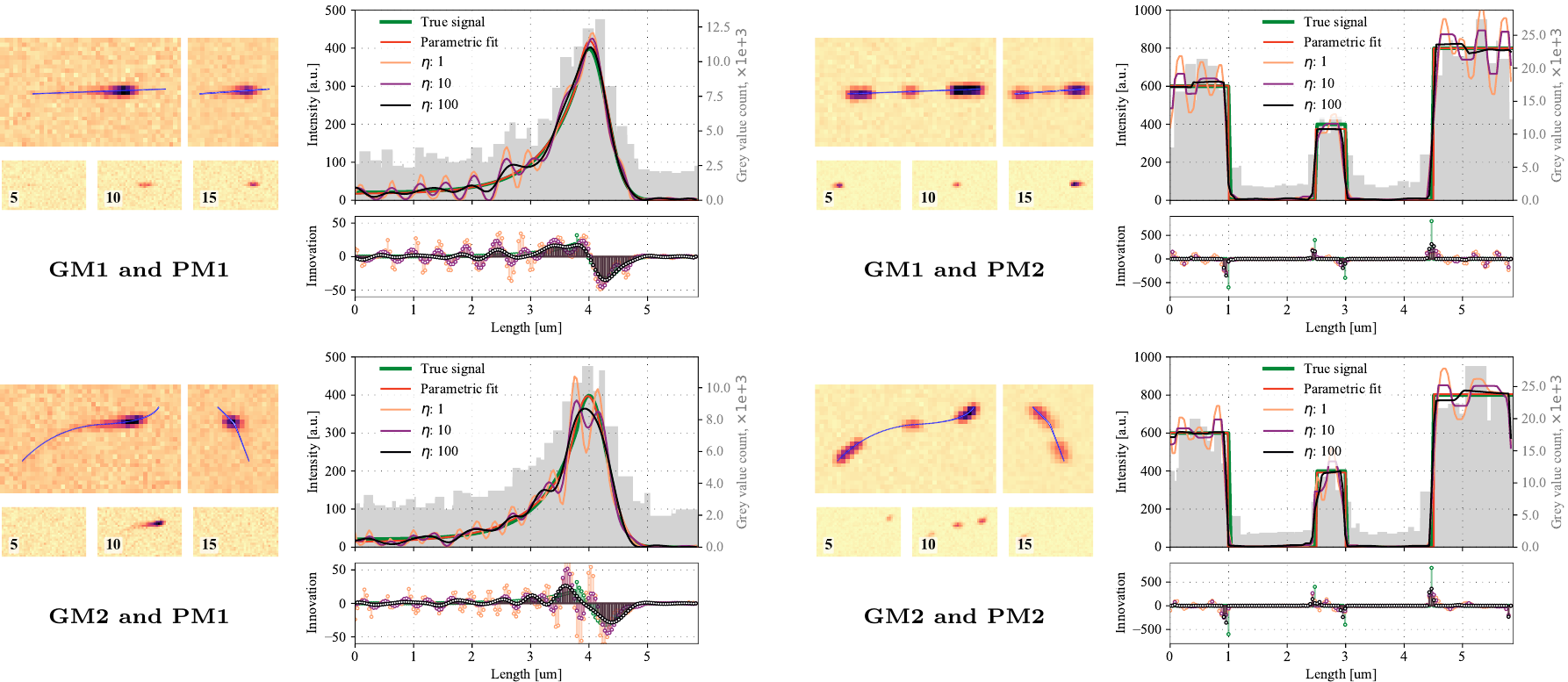}
  \caption{
    \textbf{Reconstructing light source distributions for different geometry and
      photometry scenarios.}
    We show each combination of the models in four panels.
    Each panel is organised as follows. \emph{Left, upper part}: orthogonal z-
    and x- mean projections of the synthetic image stack with the microtubule
    lattice overlaid.
    \emph{Left, lower part}: three slices acquired at different focal distances.
    \emph{Right, upper plot}: nearest neighbour estimate (scale on the right
    axis), parametric ML estimate and non-parametric MAP for three regularisation
    parameters (for both, scale on the left axis).
    \emph{Right, lower plot}: ground truth and estimated innovation process.
    The resolution in reconstruction space is \SI{0.04}{\um}.}
  \label{fig:kymo:synthetic}
\end{figure*}




\begin{figure}[!t]
  \centering
  \includegraphics[width=0.9\linewidth]{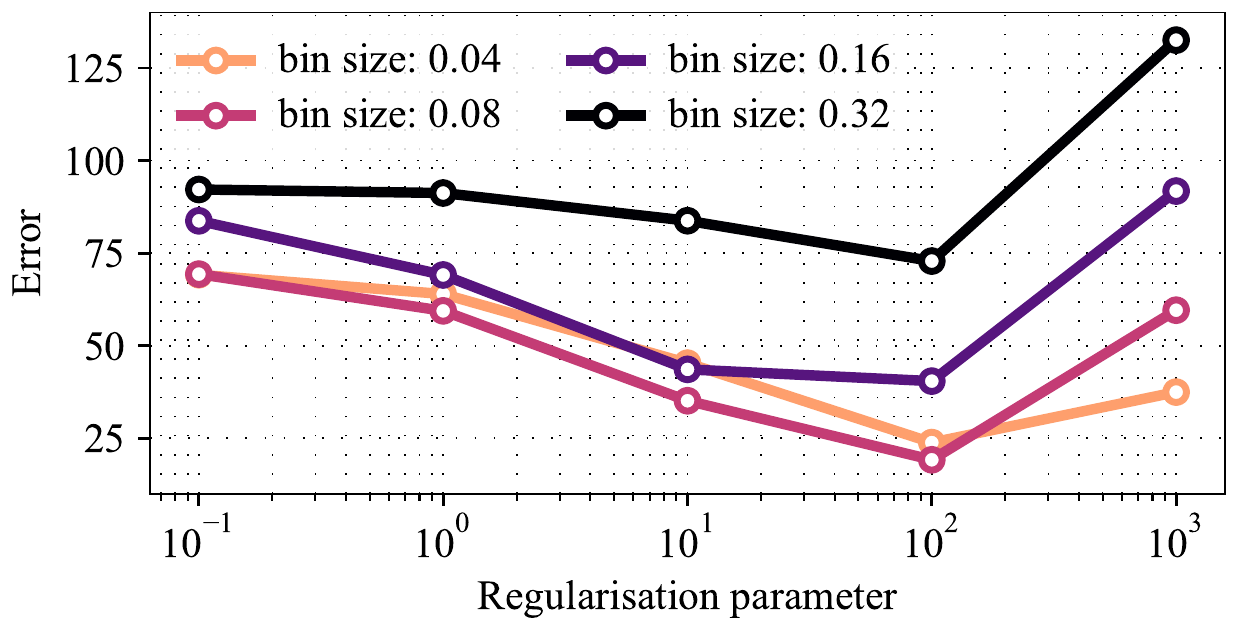}
  \caption{
    \textbf{Influence of the bin size and the regularisation parameter on the
    reconstruction accuracy.}
    The results are shown for the straight microtubule (GM1) and the
    localised fluorescence model (PM2).
    Similar trends are observed for the other combinations of geometry and
    photometry models.
    }
  \label{fig:kymo:er_vs_eq}
\end{figure}



\begin{figure}[!t]
  \centering
  \includegraphics[width=0.9\linewidth]{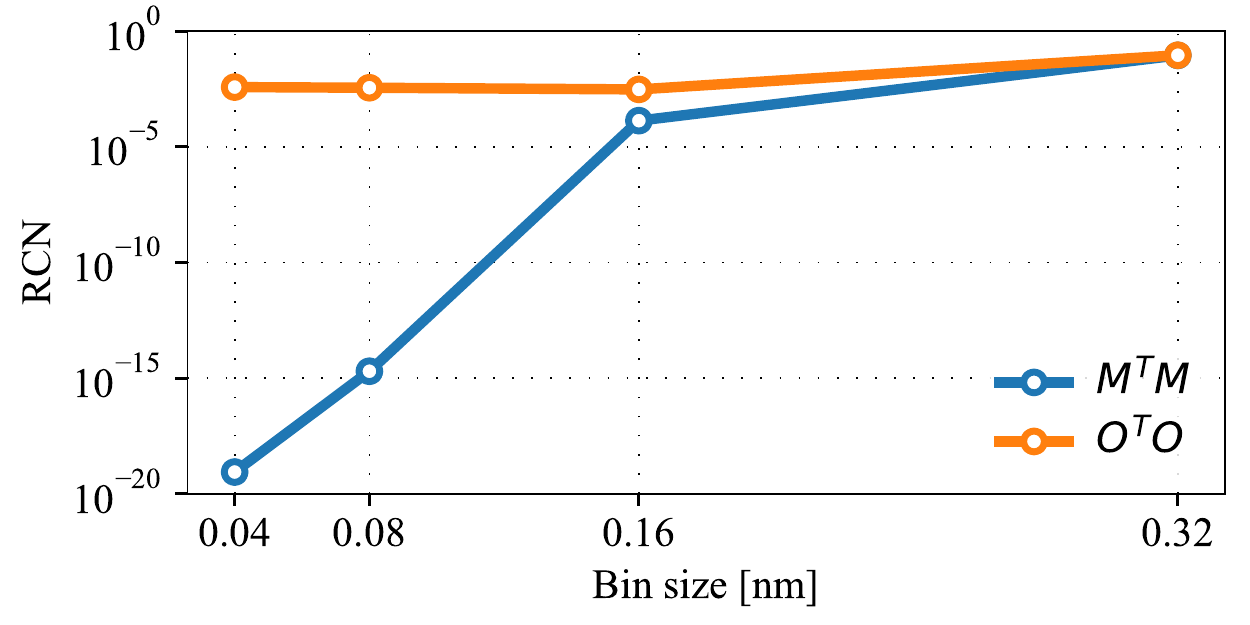}
  \caption{
    \textbf{Influence of the bin size on the reciprocal condition number.}
    The results are shown for the straight microtubule (GM1) and the
    localised fluorescence model (PM2).
    Similar trends are observed for the other combinations of geometry and
    photometry models.
    }
  \label{fig:kymo:rcn_vs_bs}
\end{figure}


We use the virtual microscope framework~\cite{Samuylov2015} to demonstrate the
capabilities of the proposed framework to reconstruct the fluorescence signal
along curvilinear objects in different geometric and photometric scenarios.
We generate a synthetic data set to compare three inverse problems: the nearest
neighbour estimate introduced in \refsection{sec:kymo:NN}, a maximum likelihood
(ML) estimate of a parametric model based on the image formation model developed
in \refsection{sec:kymo:image-formation-model}, and our MAP estimate defined in
\refsection{sec:kymo:MAP}.
In what follows, we use models and examples inspired from the microtubule
dynamics literature, but the insight applies to any curvilinear objects and is
therefore generic.

\subsubsection{Synthetic examples from microtubule models}
\label{sec:synth-exampl-from}
We consider two types of geometry models (GM), and two types of photometry
models (PM): straight and curved objects (GM1 and GM2, respectively), where
light sources are distributed smoothly (PM1, the comet shape model) or localised
in specific regions (PM2, the islands model).

\paragraph{Straight microtubule (GM1)}
We model a straight microtubule using a line segment. The mapping from the
parameter space to the physical space is defined as follows:
$\mapParamToObj \paren{\len} \defeq \psPos[o] + \curveDir \len$,
where $\psPos[o] \in \R^3$ is the origin of the microtubule, and $\curveDir \in
\R^3$ is the unit vector defining its direction. In this model, the curve is
arc-length parameterised (\ie $\RiemannianMetric = 1$), and the weighing of the
virtual sources depends only on the quadrature weights.

\paragraph{Curved microtubule (GM2)}
We model a curved microtubule using a quadratic spline function defined by four
control points sampled uniformly in the imaging volume.

\paragraph{Smooth fluorescence signal (PM1, the comet shape model)}
This photometry model uses an intensity distribution similar to the one
estimated in~\cite{Bieling2007}. In this study, it has been shown that the
protein Mal3 accumulates at the growing end of the microtubule forming a comet
shape. The observed fluorescence signal was modelled using a superposition of a
Gaussian and an exponential function described by a set of six parameters,
denoted
$\params^\pmSmooth \defeq \set{\mu, \sigma, a, b, c, d} \in \paramSet^\pmSmooth
\subset \Rpos^6$:
$\mtPhi[\params]^\pmSmooth\paren{\len}$
is defined as
$\frac{a}{\sqrt{2\pi\sigma^2}}\,\expe^{-\frac{1}{2\sigma^2}\paren{\len-\mu}^2}$,
if $\len \ge \mu - b$, and
$\paren*{\frac{a}{\sqrt{2\pi\sigma^2}}\,\expe^{-\frac{b^2}{2\sigma^2}} -
  d}\,\expe^{\frac{1}{c}\paren{\len - (\mu - b)}} + d$, else.
In~\cite{Bieling2007}, this model is fitted on the average of multiple and
aligned microtubule images. However, the precise estimated parameters were not
reported. We set the following values to visually reproduce the shape of the
photometry density: $\mu=4$, $\sigma=0.3$, $a=300$, $b=0.2$, $c=0.6$, and
$d=20$.

\paragraph{Localised fluorescence model (PM2, the island model)}
This model assumes a piecewise constant distribution of the light sources along
the curvilinear object. This choice is motivated by the fact that some
microtubule-associated proteins accumulate on specific locations on the
microtubule and appear as blurred spots (\eg \cite{Vale1996}). We define this
distribution with seven parameters, denoted
$\params^\pmSparse \defeq \set{\varphi_1, \varphi_2, \varphi_3, a, b, c, d} \in
\paramSet^\pmSparse \subset \Rpos^7$:
$\mtPhi[\params]^\pmSparse\paren{\len}$ is defined as $\varphi_1$ on
$\intervalcc{0}{a}$, $\varphi_2$ on $\intervalcc{b}{c}$, $\varphi_3$ on
$\intervalcc{d}{\mtLen}$, and $0$ else.
We fix the model parameters to the following values: $\varphi_1 = 600$,
$\varphi_2=400$, $\varphi_3=800$, $a=1.0$, $b=2.5$, $c=3.0$, $d=4.5$.

\subsubsection{Image formation model}
Using the forward model introduced in \refsection{sec:kymo:fwd}, we compute
synthetic image data for all combinations of geometry and photometry models
introduced above (\reffig{fig:kymo:synthetic}). Each image stack consists of
$21$ image slices of size $25\times42$ pixels, acquired at different focal
plans. We use a custom approximation of the microscope PSF using a Gaussian
probability density function~\cite{Zhang2007,Aguet2009}:
$\psfFunSG\paren{\x} = \constSG \, \expe^{-\frac{1}{2}\x^T\Sigma^{-1}\x}$,
where $\constSG = \paren{8\pi^3\sigma_{xy}^4\sigma_{z}^2}^{-0.5}$ is the
normalisation constant,
$\Sigma \defeq \diag\paren{\sigma_{xy}^2, \sigma_{xy}^2, \sigma_{z}^2} \in
\R^{3\times3}$ is the covariance matrix, $\sigma_{xy}$ and $\sigma_{z}$ are the
standard deviations in lateral and axial directions, respectively. The parameter
values of the image formation model are calibrated to the experimental setup
described in \refsection{seq:kymo:real-data} and summarised in
\reftable{table:params}.

\begin{table}[!t]
  \renewcommand{\arraystretch}{1.3}
  \newcommand{\head}[1]{\textnormal{\textbf{#1}}}

  \colorlet{tableheadcolor}{gray!25} 
  \newcommand{\headcol}{\rowcolor{tableheadcolor}} %
  \setlength{\aboverulesep}{0pt}
  \setlength{\belowrulesep}{0pt}

  \caption{Parameters of the image formation process.}
  \label{table:params}
  \centering
  \begin{tabularx}{0.9\linewidth}{Xcc}
    \toprule
    \headcol \head{Parameter} & \head{Notation} & \head{Value} \\
    \midrule
    Acquisition time & \tA & \SI{26.0}{\ms} \\
    Pixel size & – & \SI{160}{\nm} \\

    PSF lateral standard deviation & $\sigma_{xy}$ & \SI{130}{\nm}\\
    PSF axial standard deviation & $\sigma_z$ & \SI{255}{\nm}\\

    Quantum efficiency & $\quantEfficiency$ & $0.7$\\
    Multiplication gain & $\mGain$ & $1200$\\
    ADU conversion & $\ADU$ & $6.44$\\
    Camera bias & $\cameraBias$ & $1839$\\
    \bottomrule
  \end{tabularx}
\end{table}

\subsubsection{Comparison of the inverse problems}
\label{sec:inverse-problems}
Given the synthetic image data, we compare the fluorescent signal reconstructed
along the ground truth object geometry using three inverse problems.

\paragraph{Nearest neighbour estimate}
We apply the nearest neighbour~(NN) estimate described in
\refsection{sec:kymo:NN}. We observe in \reffig{fig:kymo:synthetic} that the
reconstruction is corrupted by bias, noise, and blur. The bias is apparent in
the low-intensity regions in the ground truth signal, where in the NN estimate
the background level remains. The effect of blurring translates in the NN
reconstruction in unsharp boundaries for the piecewise-constant photometry model
PM2. The effect of the pixel size, and the Poisson noise are also prominent in
the reconstruction. Aligning and averaging multiple image stacks as
in~\cite{Bieling2007} would alleviate the latter issues, but not the former
ones. Moreover, the NN signal is sampled directly from grey values and cannot be
easily compared with the underlying ground truth intensity signal. In
\reffig{fig:kymo:synthetic}, we display a separate axis to provide the scale of
the NN estimate. However, in the rest of the paper we omit it and compare only
the shape of NN reconstructions with other estimates.

\paragraph{Parametric ML estimate}
When the practitioner is confident in a given parametric model, one can use the
forward model described in \refsection{sec:kymo:fwd} to fit the model parameters
by minimising the negative log-likelihood, which is equivalent to maximum
likelihood. To estimate the parameters of PM1 and PM2, we solve
$\paramsHat= \arg\min_\params \nll\paren{\mtPhi[\params] \, | \,
  \obsGVMat[\iFrame]}$
using the covariance matrix adaptation evolutionary strategy (CMA-ES,
\cite{Hansen2003}). We observe that the reconstructions are close to the
underlying ground truth (\reffig{fig:kymo:synthetic}). This improves
significantly the NN reconstruction in two ways: by taking into account the
image formation model (thus reconstructing the signal in physical space), using
a single image stack (\ie without averaging many image stacks).

We interpret the NN estimate as a worst case scenario. On the other hand, the
reconstruction obtained by fitting the parametric model corresponding to the
ground truth sets a best case for the reconstruction accuracy. Nonetheless, in
real-world applications, the ground truth model is not necessarily available,
and one must rely on weaker models, such as non-parametric ones.

\paragraph{Non-parametric MAP estimate}
The smoothed photometry model PM1 can be accommodated by penalising the
squared-$\ell_2$ norm of the signal transformed using a linear operator
\cite{Bostan2013}. In the general framework of~\cite{Unser2014}, this
corresponds to choosing a Gaussian distribution to model the innovation process:
$\probInnov^{\pmSmooth}\paren{u} \propto \exp\paren{-\eta{u^2}}$
and the first-order derivative for the whitening operator.

The piecewise-constant photometry model PM2 features sparse first-order
derivatives. Therefore, we use, as for PM1, the first-order derivative as the
whitening operator, but model the innovation process using a sparsity-inducing
distribution. For example, it is possible to capture the sparsity in the
innovation process by modelling it with a Student's or Cauchy distribution.
However, in this case, the resulting optimisation problem is not convex. To
benefit from the convexity, we choose the Laplace distribution, that is often
used as a sparsity-inducing prior:
$ \probInnov^{\pmSparse}\paren{u} \propto \exp\paren{-\eta\abs{u}}$.

We set the resolution in reconstruction space to \SI{0.04}{\um}, fix the object
geometry and the background intensity to the ground truth, and use a
Gauss-Legendre quadrature with 10 points per basis function for approximating
the convolution integral~\eqref{eq:kymo:obj-mt}.
We apply \refalg{alg:full} to reconstruct the fluorescence signal for different
regularisation parameters. At a lower regularisation, we observe oscillations in
the reconstruction. However, the overall shape (in PM1) and support (in PM2) of
the signal are correctly identified. When the regularisation parameter is
increased, the amplitude of the oscillations decreases and the reconstructions
approach the ground truth.

Using a virtual microscope approach, we have shown that our framework allows
reconstructing the fluorescence signal for various scenarios of object geometry
and photometry.

\paragraph{Robustness of the MAP signal to regularisation and bin size}
The reconstruction accuracy is influenced by the regularisation parameter
(\reffig{fig:kymo:synthetic}) and the bin size.
We quantify the reconstruction error as a function of both parameters on
\reffig{fig:kymo:er_vs_eq}.
We define the error as the $\ell_1$ norm of the difference between ground truth
and reconstructions.
We observe that using higher resolution in reconstruction space results in lower
errors.
However, when the bin size is too small, the error starts increasing again.
We also observe a V-shaped trend in the error versus regularisation parameter.
Therefore, this suggests selecting an optimal regularisation, \eg using the
virtual microscope framework~\cite{Samuylov2015}, given that a model of the
underlying photometry model is available.
Remarkably, the value of the regularisation parameter resulting in the lowest
reconstruction error is the same for different bin sizes.

\paragraph{Influence of the bin size on the reciprocal condition number}
We use a constant number of virtual point sources (quadrature points) to
integrate over each bin.
If the bins are smaller, then the virtual sources are located closer to each
other and have higher overlap among the PSF supports.
As a result (see Result~\ref{result:rcn}), the RCN of the operator
$\transpose{\kymoToImageMat[\iFrame]} \kymoToImageMat[\iFrame]$ decreases.
However, when we solve the least squares sub-problem \refeq{eq:kymo:sp-ls}, we
invert the operator
$\transpose{\operator_\iFrame} \operator_\iFrame =
\transpose{\kymoToImageMat[\iFrame]} \kymoToImageMat[\iFrame] +
\transpose{\whitenedOpMat[\iFrame]} \whitenedOpMat[\iFrame] + \eyeMat_\iFrame$.
Both the regulariser (contributing
$\transpose{\whitenedOpMat[\iFrame]} \whitenedOpMat[\iFrame]$) and the
fully-splitting strategy (contributing $\eyeMat_\iFrame$) help regularising the
problem.
As a result, the conditioning of the least-squares problem improves, and gets
less sensitive to the bin size, see~\reffig{fig:kymo:rcn_vs_bs}.

\subsection{MT dynamics in \SCerevisiae}


\begin{figure*}[!t]
  \centering
  \includegraphics[width=\textwidth]{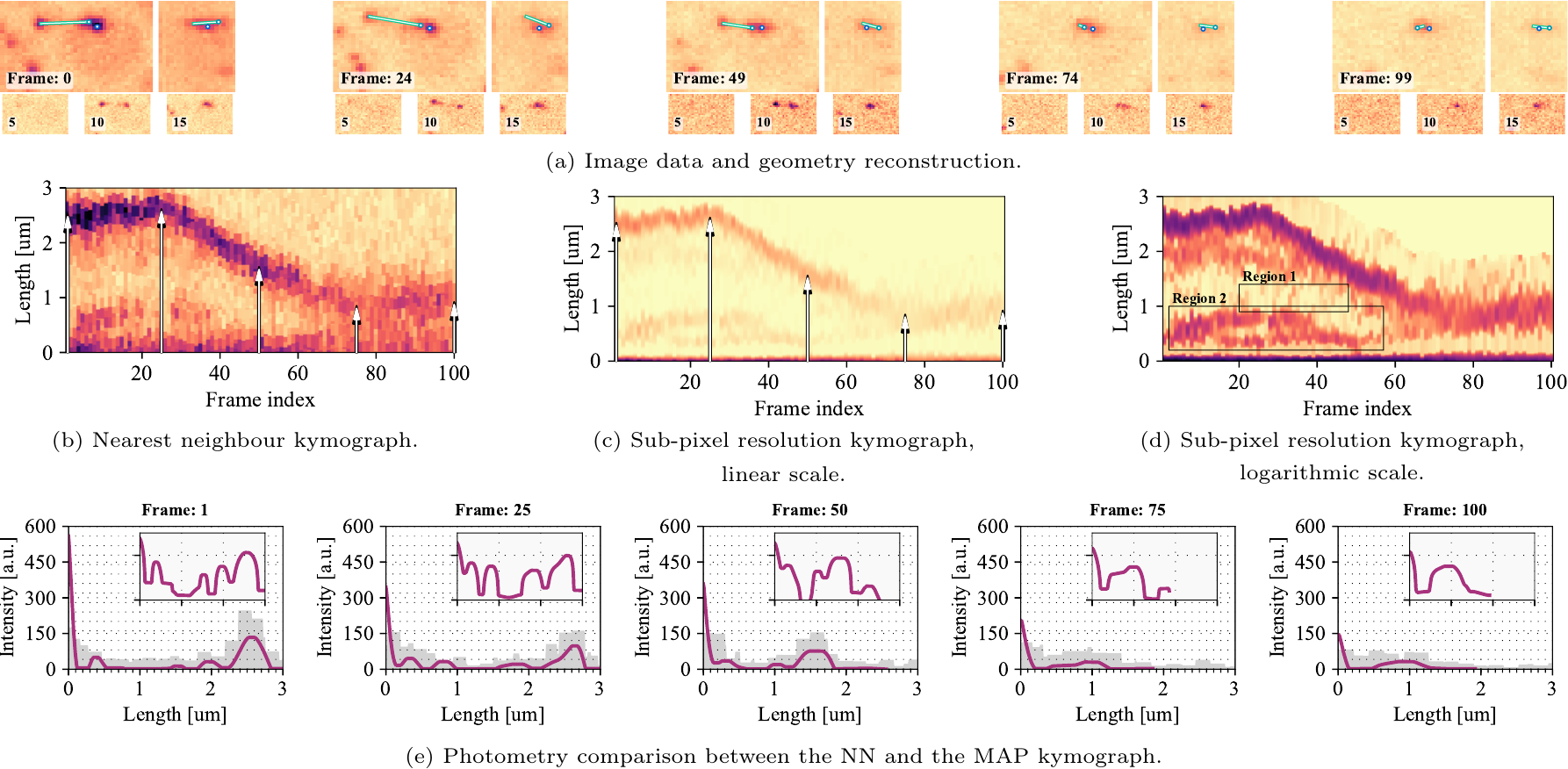}
  \subfloat{\label{fig:kymo:real:a}}
  \subfloat{\label{fig:kymo:real:b}}
  \subfloat{\label{fig:kymo:real:c}}
  \subfloat{\label{fig:kymo:real:d}}
  \subfloat{\label{fig:kymo:real:e}}
  \caption{\textbf{Sub-pixel resolution kymograph reconstruction for a single
      \SCerevisiae cell}. %
    \textit{\refsubfig{fig:kymo:real:a} Image data and geometry reconstruction}.
    Orthogonal z- and x- mean projections and image slices overlaid with the
    microtubule lattice and spindle pole bodies estimated by the particle filter
    introduced in~\cite{Samuylov2017}. Pixel size: \SI{160}{\nm}.
    \emph{\refsubfig{fig:kymo:real:b} Nearest neighbour kymograph}. Colour scale
    proportional to grey values.
    \emph{\refsubfig{fig:kymo:real:c} Sub-pixel resolution kymograph, linear scale}.
    Resolution of the reconstruction space: \SI{8}{\nm}. Colour scale
    proportional to photon counts.
    In \emph{\refsubfig{fig:kymo:real:b}} and
    \emph{\refsubfig{fig:kymo:real:c}}, the microtubule geometry estimated by the
    particle filter is overlaid as an upward arrow. The plus-end and the minus-end
    are located at the head and tail respectively.
    \emph{\refsubfig{fig:kymo:real:d} Sub-pixel resolution kymograph,
      logarithmic scale}. Logarithmic colour scale shown for a better contrast. We
    observe patterns of dynamics of small clusters at a high rate (\eg within
    \emph{region 1}) and of large clusters at a slower rate (\eg within \emph{region
      2}).
    \emph{\refsubfig{fig:kymo:real:e} Photometry comparison between the NN and
      the MAP kymograph}.
    The photometry along the arrows displayed in
    \emph{\refsubfig{fig:kymo:real:b}} and \emph{\refsubfig{fig:kymo:real:c}} is
    shown: NN estimate shown as a grey histogram; MAP estimate shown as a purple
    curve; Logarithmic signal shown in the inset).
    }
  \label{fig:kymo:real}
\end{figure*}



\begin{figure*}[!t]
  \centering
  \includegraphics[width=\textwidth]{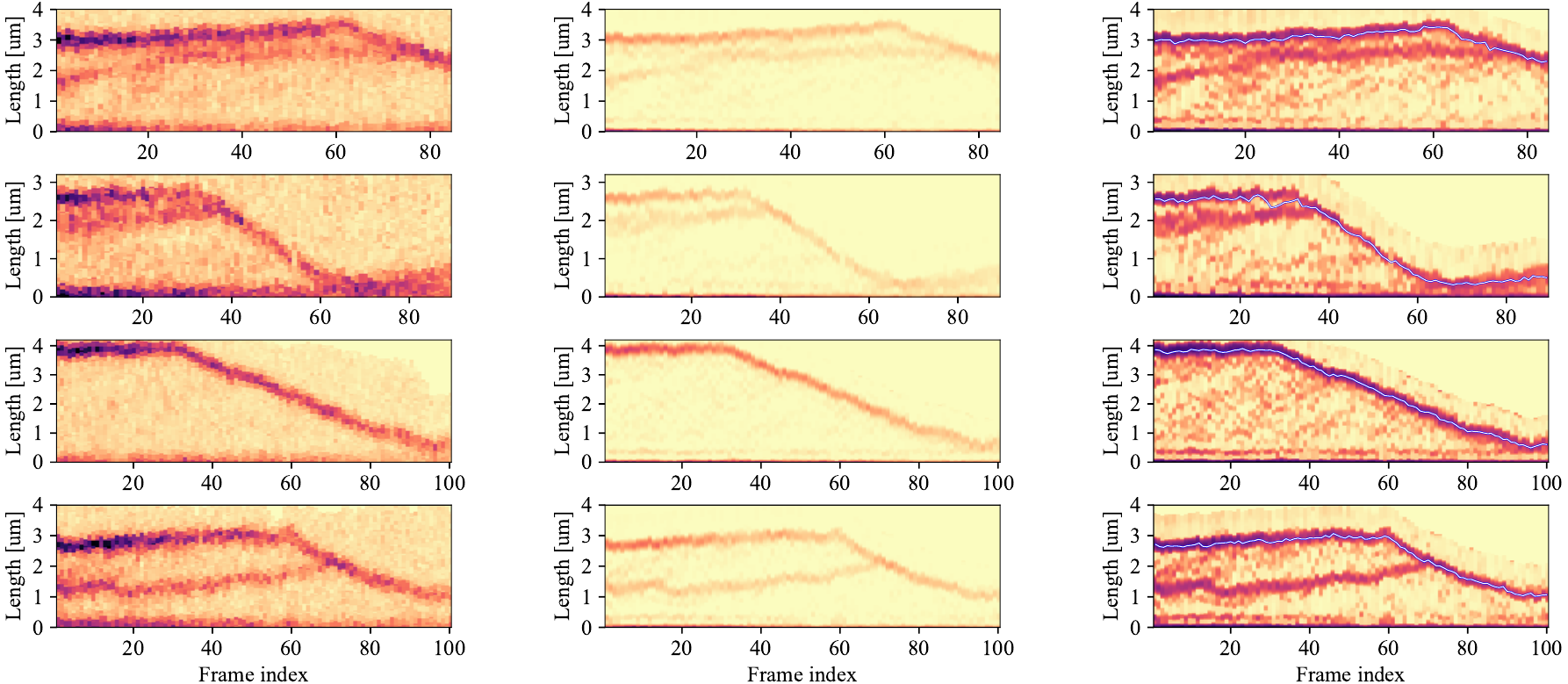}
  \caption{
    \textbf{Diversity of the pattern of dynamics in the \SCerevisiae \invivo
      kymographs}.
    Each row corresponds to a different \SCerevisiae cell.
    \emph{First column}: NN kymograph, colour scale proportional to grey values.
    \emph{Second column}: MAP kymograph, linear colour scale proportional to
    photon count.
    \emph{Third column}: MAP kymograph, logarithmic colour scale for better
    contrast.
    The microtubule length estimated by the particle filter is overlaid on the
    kymographs as a white curve.
    We observe that the estimator is biased towards the maximum of the
    comet-shaped intensity signal at the microtubule plus-end, and it results in an
    underestimated microtubule length.}
  \label{fig:kymo:gallery}
\end{figure*}


In this section we apply our framework to reconstruct sub-pixel resolution
kymographs of microtubule dynamics from time-lapse images acquired \invivo in
the budding yeast~\cite{Rauch2010}.

\subsubsection{MT dynamics image data}
We show the results for the strain $\mathrm{Kip3}\Delta$, imaged in
pre-anaphase. This dataset contains 743 cells. The images are characterised by
the presence of a single straight cytoplasmic microtubule that is longer than
the one observed in wild type yeast cells.
The microtubule ends are visualised using two families of proteins labeled with
the green fluorescent protein: Spc72p binding specifically to the spindle pole
bodies (structures from where the microtubule minus-end originates), and Bik1
accumulating at the microtubule plus-end~\cite{Rauch2010}.
We observe that most of the fluorescence signal is located at the microtubule
ends (\reffig{fig:kymo:real:a}). However, the protein Bik1 can attach to and be
transported along the microtubule lattice~\cite{Rauch2010}.

We assume that the distribution of light sources along the microtubule is
sparse. Therefore, we model the underlying intensity signal using the first
derivative as the whitening operator and assume that the increments of the
innovation process follow a Laplace distribution.

\subsubsection{Imaging parameters}
\label{seq:kymo:real-data}
The images are acquired by a spinning disk confocal microscope equipped with a
63X 1.4 NA objective, \SI{493}{\nm} solid-state laser, and EM-CCD camera
(Hamamatsu ImageEM).
An image stack of size $21\times256\times256$ pixels is acquired every $0.55$
seconds, resulting in a sequence of $100$ frames.
The regions of interest containing individual cells in pre-anaphase are manually
outlined and cropped out. The imaging parameters are summarised in
\reftable{table:params}.

\subsubsection{Reconstructing the background intensity}
To estimate the photobleaching of the background intensity, we average the grey
values of the pixels in each image stack, convert them into photon counts by
applying the inverse of the pixel--to--image mapping $\mapPxToImg$ given in
\refeq{eq:kymo:pixel-to-image-map}, dividing the result by the spatio-temporal
integration constant given by $\pxArea \tA$ in
\refeq{eq:kymo:ifm-general-integrated}. We fit an exponential decay model
(characterised by three parameters: an amplitude, a decay rate and an offset). We
minimise the $\ell_2$ norm of the residuals using the CMA-ES algorithm
\cite{Hansen2003}.

\subsubsection{Reconstructing the microtubule geometry}
\label{seq:kymo:real-data:geom}
Reconstructing kymographs at sub-pixel resolution using our framework requires
first to reconstruct the microtubule geometry at each frame. Several frameworks
exist to track microtubule ends in fluorescence microscopy image data (\eg
\cite{Smal2008,Cardinale2009,Matov2010,Samuylov2017}). We use the particle
filter strategy introduced in~\cite{Samuylov2017}. It is designed to track
microtubules in \SCerevisiae. It relies on models that explicitly encode the
microtubule lattice geometry but only accounts for the intensity signal at its
ends, because this is the original idea behind the genetic engineering
introduced in~\cite{Rauch2010}. We note that it is a special case of the object
model presented in this paper. Modelling the fluorescence signal as a linear
combination of two Dirac measures with atoms at the ends of the microtubule
lattice, we recover from \refeq{eq:kymo:object-model-curve} the microtubule
model from~\cite{Samuylov2017}:
\begin{equation}
  \label{eq:kymo:object-model-curve-ps}
  \objMeas^{\C\,\PS}\paren{\dy\times\dt} =
  \1{\curve}\paren{\y} \sum_{\mathclap{\len \in \set{0, \, \curveLen}}}
  \mtPhi\paren{\len} \dirac_{\mapParamToObj \paren{\len}}\paren{\dy} \dt \ .
\end{equation}
The tracking results for one cell are shown in \reffig{fig:kymo:real:a}. We visualise
the spindle pole bodies as two points and the microtubule lattice as a line
segment.

\subsubsection{Reconstructing sub-pixel resolution kymographs}
We set the reconstruction space resolution to \SI{8}{\nm}. This is comparable
with the size of a tubulin dimer, the building block of
microtubules~\cite{Akhmanova2008}. The microtubule length estimated using the
particle filter is extended up to \SI{3}{\um} for the NN reconstruction and to
\SI{1}{\um} for our framework.
The regularisation parameter is set to $\reg = 1$.
\reffig{fig:kymo:real} shows the results for a single cell, comparing the NN
reconstruction with our framework.

The NN kymograph suffers from a coarser resolution, higher level of noise and a
higher impact of blurring (\reffig{fig:kymo:real:b}). The fluorescence signal at the
microtubule \emph{minus-end} (bottom of the kymograph) appears to be weaker than
at the \emph{plus-end} (top of the kymograph).
The sub-pixel resolution kymograph (\reffig{fig:kymo:real:c}), appears
qualitatively different. The signal at the minus-end is sharply peaked, and its
width is about \SI{200}{\nm}. It is consistent with the fact that the protein
Spc72p is used as a minus-end marker: it is known to be anchored in the outer
plaque of the spindle pole bodies, which have a width of about \SI{185}{\nm}
\cite{Rauch2010,Bullitt1997}. The plus-end signal has a lower amplitude than at
the minus-end, and it shows a comet-like shape similar to the pattern of the
end-binding protein Mal3 observed \invitro~\cite{Bieling2007}.
In addition, our framework allows revealing dynamical patterns that are not
observed \invitro by averaging multiple microtubules~\cite{Bieling2007}, and
goes beyond the two-point photometry hypothesis used in
\cite{Cardinale2009,Samuylov2017}. Indeed, we observe patterns along the
microtubule lattice (\reffig{fig:kymo:real:d}). This signal is weaker than at
the microtubule ends, but still significant. We can distinguish the dynamics of
small clusters at a high rate (\eg within region 1) and of large clusters at a
slower rate (\eg within the region 2), displaying merging and splitting events.
Most of these events cannot be observed as clearly from the NN reconstruction
(\reffig{fig:kymo:real:e}).

The advantage of formulating kymographs as an inverse problem accounting for
optical distortions, noise, and sampling effects allows reconstructions based on
individual microtubule image data. This is crucial for studying microtubule
dynamics where the stochasticity is built-in and generate a large diversity of
patterns, as shown in \reffig{fig:kymo:gallery}.

\subsubsection{Validating  sub-pixel resolution kymographs}

\begin{figure*}[!t]
  \centering
  \includegraphics[width=\textwidth]{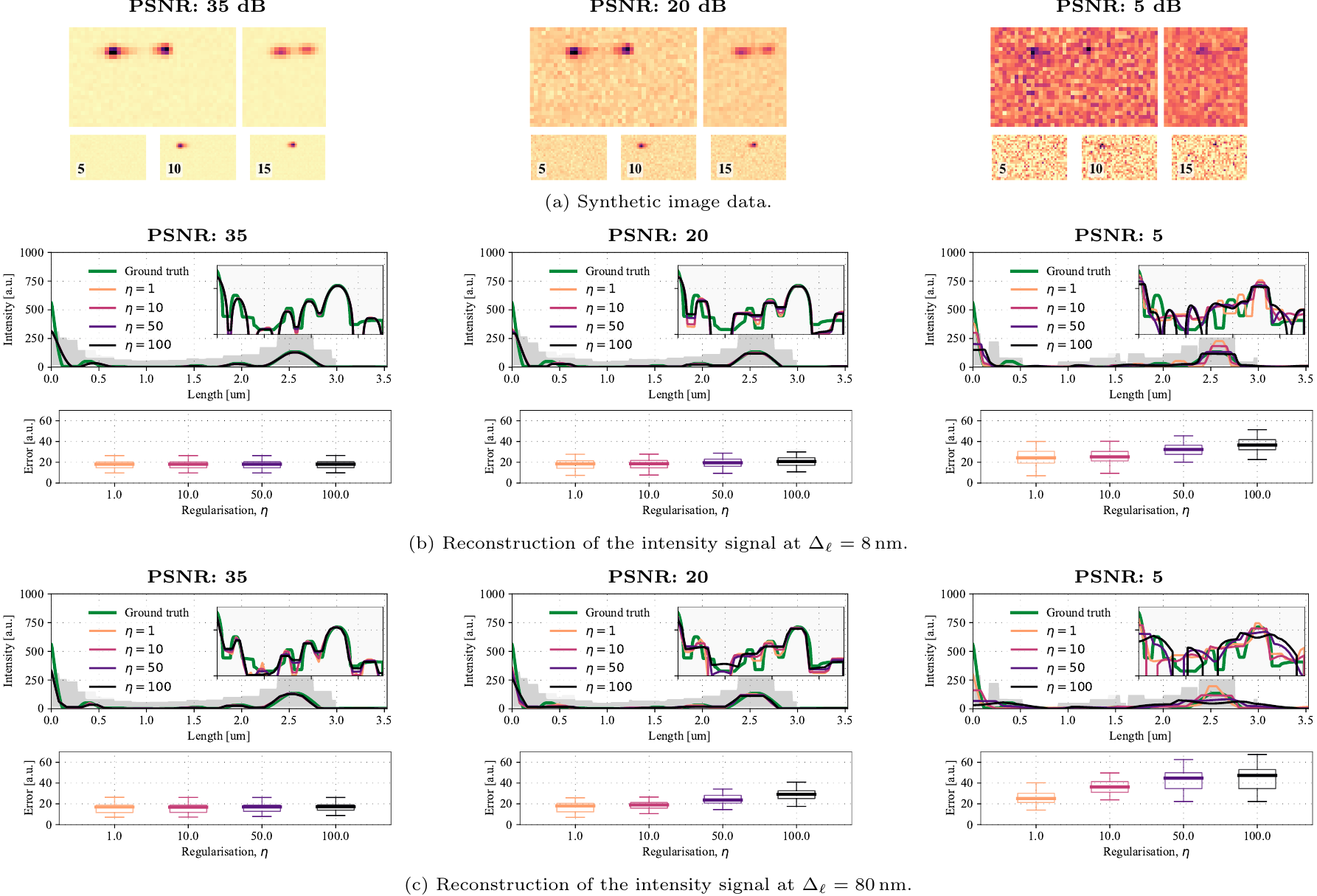}
  \subfloat{\label{fig:kymo:validation:a}}
  \subfloat{\label{fig:kymo:validation:b}}
  \subfloat{\label{fig:kymo:validation:c}}
  \caption{
    \textbf{Robustness of the MAP kymograph to \psnr, regularisation ($\reg$),
    and bin size ($\binSize$)}.
    \emph{\refsubfig{fig:kymo:validation:a} Image data}. Orthogonal z- and x-
    mean projections and image slices computed using the virtual microscope framework at different PSNR.
    The ground truth is the MAP estimate shown in
    \emph{\reffig{fig:kymo:real}}~\emph{\refsubfig{fig:kymo:real:c}}.
    \emph{Robustness to \psnr and regularisation for two bin sizes: $\binSize =
      \SI{8}{\nm}$ \refsubfig{fig:kymo:validation:b}, and $\binSize = \SI{80}{\nm}$
      \refsubfig{fig:kymo:validation:c}}.
    Each panel is organised as follows.
    \emph{First row}: For each bin size, we compare the ground truth profile for
    the first frame with the NN estimate and the MAP estimate.
    The photometry reconstruction in a logarithmic scale are shown as an inset.
    \emph{Second row}: box plot of the per-frame reconstruction error.
    The reconstruction error is the root mean squared error between the
    estimated and the ground truth profile.
  }
  \label{fig:kymo:validation}
\end{figure*}



%
We use the sub-pixel reconstruction shown in \reffig{fig:kymo:real} as the
ground truth. We simulate data according to the model described in
\refsection{sec:kymo:fwd} at different peak signal-to-noise ratio (\psnr,
adapted from \cite{Aguet2009}), defined as:
\begin{equation*}
  \tA = \frac{
  \bgPhi[0] + \frac{1}{\nPx} \sum_{\iPx\in\iPxSet}{ \phFlux[\C]\paren{\x_\iPx, 0} }
  }{
  \pxArea \max_{ \iPx\in\iPxSet }\paren*{ \phFlux[\C]\paren{\x_\iPx, 0} }^2
  } \,
  10^{0.1\,\psnr} \ ,
\end{equation*}
where $\phFlux[\C]\paren{\x_\iPx, 0}$ is evaluated using the virtual source
representation of \refeq{eq:kymo:ph-flux-c}.

We use this synthetic dataset to investigate two important factors: the
regularisation parameter and the spatial resolution of the reconstruction space
(see \reffig{fig:kymo:validation:b} and \reffig{fig:kymo:validation:c}).
We observe that at high \psnr the signal is robust to changes in both the
regularisation and the resolution. However, a higher resolution of the
reconstruction space yields a more accurate reconstruction of the signal at a
lower \psnr.

\subsubsection{Length bias induced by a simplified photometry model}
In \reffig{fig:kymo:gallery}, the microtubule length estimated by the particle
filter is displayed on top of the kymographs. We observe that the microtubule
length is always underestimated. As discussed in
\refsection{seq:kymo:real-data:geom}, the tracker models the fluorescence
distribution with two point sources located at the microtubule ends. However, as
observed in the kymographs, the fluorescence distribution at the microtubule
plus-end has a comet-like shape. Thus, the point source approximation of the
photometry is biased. As a result, the particle filter estimates the microtubule
plus-end at the local maximum of the comet. From the reconstructions we observe
that the size of the comet changes in time, and the bias in the microtubule
length using this simplified photometry model is variable. Finally, the particle
filter can drift to a large cluster near the plus-end. This results in a
spurious jump in the microtubule length dynamics (see \reffig{fig:kymo:gallery},
second row). Therefore, there is a coupling between geometry and photometry
estimation: incorrect assumptions about one of affects the reconstruction of the
other.




\section{Discussion}

In the existing literature, the degradations entailed by the image formation
process are handled either before or after the kymograph reconstruction.
One approach is to account for the degradations by denoising and deconvolving
the original image as a pre-processing step~\cite{Racine2007,Kner2009}.
However, standard restoration algorithms are implicitly based on models that are
not designed for singular objects, such as points or curvilinear objects. For
example, an interesting direction could be to account for more general noise
models, such as mixed Poisson-Gaussian by using techniques tailored for the low
photon count regime~\cite{Makitalo2011,Luisier2011} as a pre-processing step.
Nonetheless, these techniques are designed and assessed in the digital domain,
and it is yet unclear how this will influence the kymograph reconstruction
quality. An interesting line of investigation could be to integrate these ideas
directly in our Bayesian framework.
To achieve a sub-pixel resolution of the particles identified in kymograph
space, \cite{Zhang2011} proposes to refine their position in physical space by
fitting the PSF to the original image data. However, this step is performed
\emph{a posteriori} and only for individual particles.
Another approach is to post-process the kymograph by applying digital filters in
kymograph space~\cite{Racine2007,Mukherjee2010,Chetta2011}. However, this kind
of restoration approach is necessarily heuristic. Indeed, the nonlinear
embedding of the kymograph space into physical space renders the interpretation
of the signal in kymospace difficult.
To achieve only denoising, it is custom to average independent image data before
the kymograph reconstruction~\cite{Bieling2007}. However, this is meaningful
only for images of comparable objects that can be registered. Beyond this
fundamental limitation, such a procedure does not allow assessing the
inter-object variability, and will only capture the most prominent features of
the kymograph.

Our kymograph reconstruction is based on a well-grounded Bayesian framework.
However, in this paper we focus on establishing a minimal version of the
framework that can handle an analog reconstruction of the kymograph, at the
expense of several simplifications in both the forward and inverse problem.
For the forward problem, the main extension could be alleviating the
over-simplified Gaussian approximation of the PSF model.
We have shown that the PSF shape is crucial for reconstructing the photometry of
punctual objects, and it would probably also improve the photometry
reconstruction of curvilinear objects~\cite{Samuylov2017,Samuylov2017psf}.
In \cite{Samuylov2017psf}, we introduce a framework to model any PSF as a sparse
mixture of Gaussians: it provides an accurate representation of any PSF shape,
still computationally efficient to evaluate with IFGT, and compatible with the
virtual microscope framework. Therefore, such an extension would require only
minor modifications to our kymograph reconstruction framework.
Another direction to extend the forward problem relates to the pixel--to--image
mapping. In this work, we have used a deterministic affine camera mode, but more
advanced models account for the noise process in the pixel--to--image mapping
\cite{Hirsch2013,Konnik2014}. These camera models can be extended to handle a
specific type of camera. For example, pixel-dependent noise models are crucial
for the processing of images acquired using a complementary metal-oxide
semiconductor (CMOS) camera~\cite{Huang2013}.

For the inverse problem, a straightforward extension is to consider more general
sparse stochastic processes~\cite{Unser2014}. The innovation process modelling
involves a whitening operator, and a probability distribution.
We restrict ourselves to first order whitening operators. This limitation allows
simplifying the inverse problem to a trivial factorisation of the prior
probability into its marginals. For higher-order operators, the discrete
innovation process in \refeq{eq:kymo:innov-model-discrete} follows a Markov
chain with an order depending on the whitening operator order. This entails a
more involved factorisation resulting in a more complicated MAP
algorithm~\cite{Amini2013}. Nonetheless, this extension is solely based
on~\cite{Amini2013} and is straightforward to integrate in the proposed
kymograph reconstruction framework.
We also restrict ourselves to Gaussian and Laplacian distributions. The family
of sparse stochastic processes \cite{Unser2014} contains other compatible
distributions with even better sparsity-inducing capabilities. In our
fully-split formulation, this would amount to changing only the innovation
potential sub-problem \refeq{eq:kymo:sp-innov}. This is straightforward to do
for distributions having a known proximal operator~\cite{Bostan2013}. However,
the resulting minimisation problem would become nonconvex, and hence more
challenging in practice. This additional complexity should be justified by an
application for which the simpler alternatives are irrelevant.

Finally, we make two assumptions regarding independence that would require
significantly more efforts to alleviate.
We treat frames independently: it allows processing each image stack separately,
at the expense of simplifying the dynamical model taking place in kymospace.
Going beyond this limitation requires modelling the spatio-temporal dynamics of
the fluorescence signal along the geometry. A straightforward approach would be
to extend the class of priors to spatio-temporal sparse stochastic processes. We
believe that this exciting line of research will benefit from building
spatio-temporal priors based on the relevant biology, biochemistry and
biophysics literature (\eg~\cite{Reese2011}).
We also assume the geometry to be known or already estimated. The joint
reconstruction of the geometry and photometry is a problem already tackled for
punctual objects~\cite{Samuylov2017,Samuylov2017psf}. However, it is a more
challenging task for curvilinear objects because of the more complex geometry
and the feedback between the two tasks. Nevertheless, we believe that our
framework could be used as a modelling/algorithmic building block of such a more
advanced inverse problem.




\section{Conclusion}

In this work, we propose a Bayesian framework for the kymograph reconstruction
given the geometry of a curvilinear object.
This allows a well-grounded formulation of the inverse problem that, first,
involves a \emph{realistic image formation model} accounting for optical
distortions and measurement noise, second, relies on a proper reconstruction of
the \emph{fluorescence signal in physical space} modelled using a class of
flexible non-parametric priors derived from \emph{Lévy innovation processes}.
Due to the singular nature of curvilinear geometries, the kymograph
reconstruction problem is inherently \emph{analog}.
However, using the \emph{virtual microscope framework}, we formulate a
computationally tractable approximation that allows deriving efficient iterative
algorithms based on a fully-split \emph{alternating split Bregman} algorithm.

Using the virtual microscope framework, we assessed our Bayesian framework on
synthetic and real image data.
We showed that our framework allows modelling different combinations of geometry
(straight/curved) and photometry (smooth/piecewise-constant) in a unified
fashion, demonstrating the genericity of our approach. In addition, our
framework is based on an analog reconstruction space (\ie the kymograph space)
that allows interpreting the kymograph directly in terms of light source
distribution dynamics in physical space, where the signal is deblurred and
denoised.
We applied our framework to the problem of characterising microtubule dynamics
\invivo in the budding yeast \SCerevisiae. We demonstrated that the common point
source approximation of the photometry is oversimplified and that it introduces
a bias in the estimated microtubule length.
Moreover, we show that our framework allows revealing complex patterns occurring
on the microtubule lattice from single time-lapse data. These patterns are not
clearly identified with a canonical approach, such as the nearest neighbour
kymograph.

We expect that the framework proposed in this paper will facilitate the analysis
of kymographs and enable new discoveries thanks to the increase in quality and
resolution.
In addition, the framework is modular and lays the ground for extensions that
could better fit particular applications.
It will allow formulating more specific models, widening the scope of hypotheses
that can be tested in fundamental fields such as cell biology.



\section*{Acknowledgment}
\addcontentsline{toc}{section}{Acknowledgment}

We thank A. M. Rauch and Y. Barral for sharing the original yeast data. We are
also grateful for the feedback of the anonymous reviewers that greatly improved
the manuscript. This work has been supported by the SystemsX.ch RTD Grant
\#2012/192 TubeX of the Swiss National Science Foundation.

\bibliographystyle{IEEEtran}
\bibliography{IEEEabrv,bib-kymo/refs,bib-kymo/books,bib-kymo/revision}

\begin{IEEEbiography}[{\includegraphics[width=1in,height=1.25in,clip,keepaspectratio]{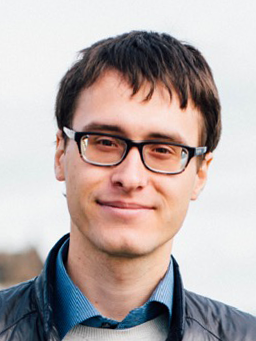}}]{Denis
    K. Samuylov}
  received the B.Sc. degree in telecommunications from the Saint Petersburg
  State Polytechnic University (currently the Peter the Great Saint Petersburg
  Polytechnic University), Russia, in 2011. He received the M.Sc. degree in
  interdisciplinary approaches to life science from the Paris Diderot University
  (Paris 7), France, in 2013. He received the Ph.D. degree in computer vision
  from the Computer Vision laboratory at the Swiss Federal Institute of
  Technology in Zurich (ETH Zurich), Switzerland, in 2018. His research
  interests include the application of signal processing and computer vision
  techniques to various problems in biology and medicine.
\end{IEEEbiography}

\begin{IEEEbiography}[{\includegraphics[width=1in,height=1.25in,clip,keepaspectratio]{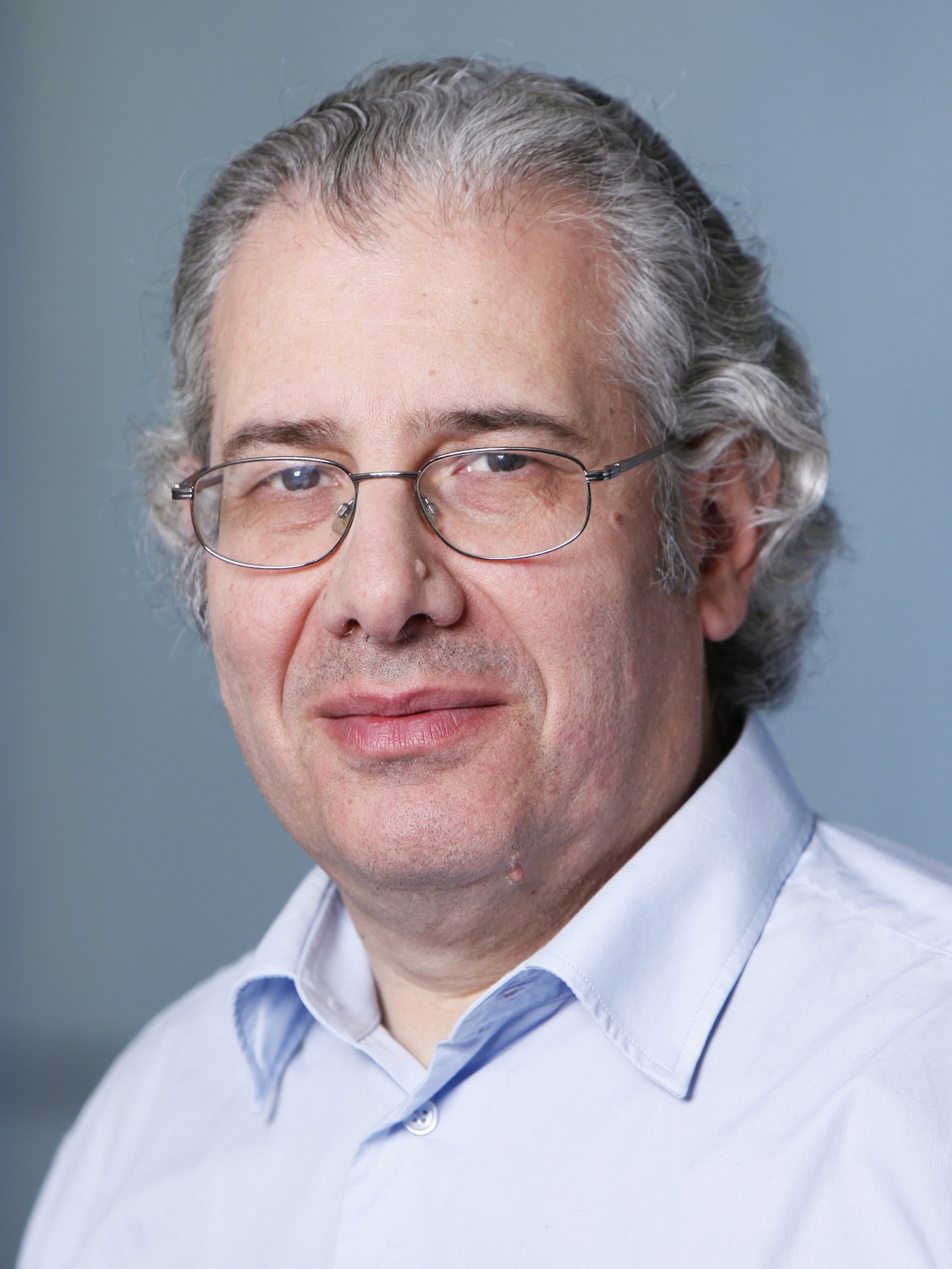}}]{Gábor
    Székely}
  received the Graduate degree in chemical engineering, the Graduate degree in
  applied mathematics, and the Ph.D. degree in analytical chemistry from the
  Technical University of Budapest and the Eötvös Lórand University, Budapest,
  Hungary, in 1974, 1981, and 1985, respectively. Since 2002 he has been leading
  the Medical Image Analysis and Visualization Group at the Computer Vision
  Laboratory of Swiss Federal Institute of Technology (ETH) Zurich, Switzerland,
  concentrating on the development of image analysis, visualization, and
  simulation methods for computer support of biomedical research, clinical
  diagnosis, therapy, training and education.
\end{IEEEbiography}

\begin{IEEEbiography}[{\includegraphics[width=1in,height=1.25in,clip,keepaspectratio]{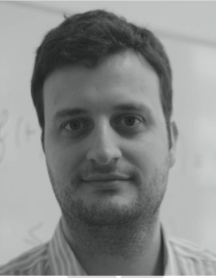}}]{Grégory
    Paul}
  received the M.Sc. degree in cell biology and physiology from Ecole Normale
  Supérieure, Paris, France, in 2003, and the Ph.D. degree from the University
  of Paris VI, in 2008. In 2008 he joined the Swiss Federal Institute of
  Technology (ETH) Zurich, Switzerland, as a post-doctoral researcher in the
  computer science department with Prof. Ivo F. Sbalzarini to develop new
  quantitative tools to investigate biological processes from image data.
  Together with Prof. Gábor Székely, between 2012 and 2017 he was leading the
  BioimagE Analysis and Modeling (BEAM) team, at ETH in the Computer Vision
  Laboratory, focusing on the development of image analysis, computational
  statistics and biophysical modeling applied to cell biology.
\end{IEEEbiography}

\vfill

\end{document}